\pdfoutput=1
\documentclass[twocolumn,prd,floatfix,nofootinbib,longbibliography,notitlepage,superscriptaddress,aps]{revtex4-2}

\usepackage{amsmath}
\usepackage{amssymb}
\usepackage{bm}
\usepackage{subfigure}
\usepackage[usenames,svgnames,dvipsnames]{xcolor}
\usepackage{array}
\usepackage{bbm}
\usepackage{float}

\usepackage{graphicx}

\graphicspath{{./Figures/}}

\usepackage[T1]{fontenc}
\usepackage[utf8]{inputenc}
\usepackage{lmodern}

\newcommand{\secref}[1]{section~\ref{#1}}
\newcommand{\subsecref}[1]{subsection~\ref{#1}}
\newcommand{\Figref}[1]{Figure~\ref{#1}}
\newcommand{\eg}{e.g.,\,}
\newcommand{\ie}{i.e.,\,}

\usepackage[unicode, colorlinks, allcolors=blue!70!black, linktocpage, pdfusetitle]{hyperref}
\hypersetup{
    colorlinks=true,
    urlcolor=SteelBlue,
    linkcolor=red,
    citecolor=blue,
}
\usepackage[all]{hypcap}
\usepackage{orcidlink}
\usepackage{microtype}

\newcommand{\trajback}{%
\begin{figure*}[t!]
\centering
\includegraphics{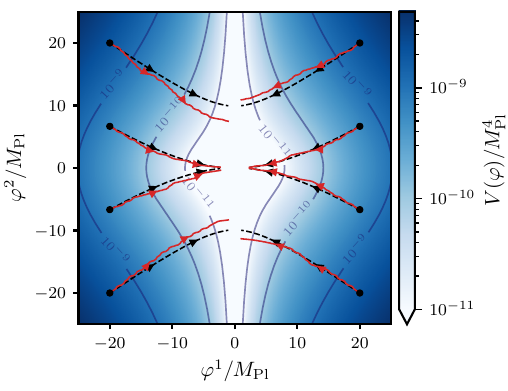}
\hfill
\includegraphics{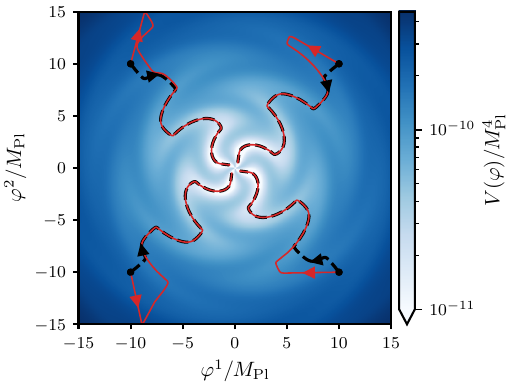}
\caption{\label{fig:def_geom_V_traj} Background field trajectories in two-field inflationary models, illustrating the effects of field-space geometries and structures in the potential.  
Left panel: Field trajectories for the nonlinear smooth potential 
$V(\varphi^1,\varphi^2)=\frac{\lambda}{4}(\varphi^1)^4+\frac{g}{2}(\varphi^2)^2(\varphi^1)^2$, deformed by curvature effects arising from a nontrivial metric (see \eqref{eq:fld_met_ex}). The  trajectories in solid red illustrate how field-space geometry induces very sharp turns in the trajectories, leading to features in the primordial curvature spectrum. For comparison, the curve in dashed black lines displays the background trajectory in flat field-space. Right panel: Field trajectories exhibiting sharp turns generated by the scalar potential in \eqref{eq:potential_st}. The solid red curves consider initial conditions in the high-velocity regime in which trajectories are not constrained to move along the potential minimum, while the trajectories in dashed black lines correspond to trajectories with zero initial velocity.}
\end{figure*}
}

\newcommand{\srparams}{%
\begin{figure*}[t!]
\centering
\includegraphics{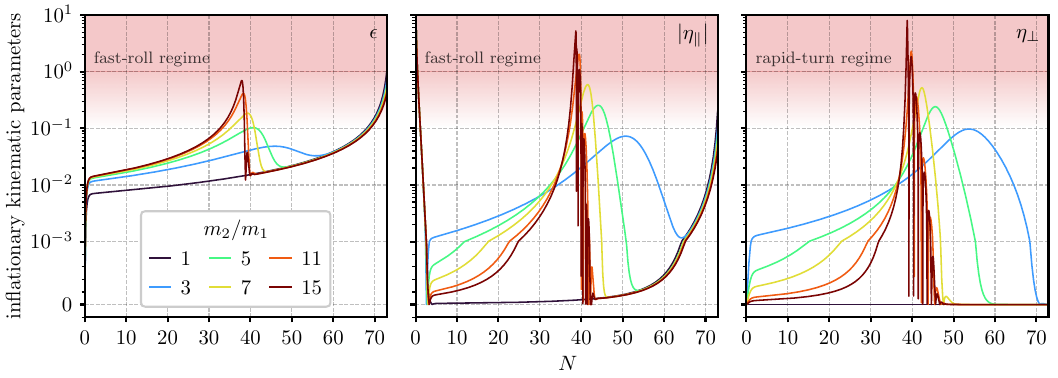}
\caption{\label{fig:slow_roll_params} Evolution of $\epsilon$ (left), $|\eta_{\parallel}|$ (center), and $\eta_{\perp}$ (right) as functions of the number of e-folds $N$, for the two-field quadratic potential with different mass ratios $m_{2}/m_{1}$. Regions where the kinematic parameters are of $\mathcal{O}(1)$ are shaded in graded tones of red. The azimuthally symmetric case $m_{2}/m_{1}=1$ gives $\eta_{\perp} \equiv 0$; increasing $m_{2}/m_{1}$ sharpens the bend of the trajectory towards the shallow valley, producing progressively narrower and taller peaks in all three quantities at the turn.}
\end{figure*}
}

\newcommand{\CDfail}{%
\begin{figure}[t!]
\centering
\includegraphics{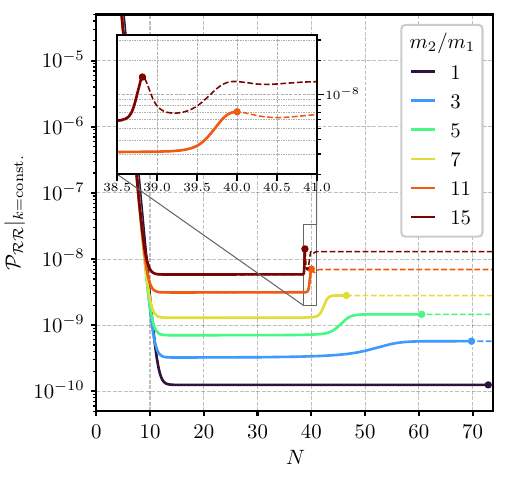}
\caption{\label{fig:cholesky_problem} Evolution of constant-$k$ curvature modes for different values of the mass ratio $m_2/m_1$. Solid lines show the evolution obtained using the Cholesky decomposition scheme presented in \cite{GalvezGhersi:2016wbu}, with 
dots at the end of the solid lines marking the point at which the 
evolution breaks down. Dashed lines depict the full mode evolution obtained 
using the evolution scheme developed in this paper.
Instabilities manifest in the Cholesky scheme as interruptions in the evolution following sharp turns in the background trajectories. The inset in the upper-left corner shows that, as the mass ratio grows, the post-turn mode evolution becomes increasingly nontrivial.}
\end{figure}
}

\newcommand{\injscheme}{%
\begin{figure}[t!]
\centering
\includegraphics{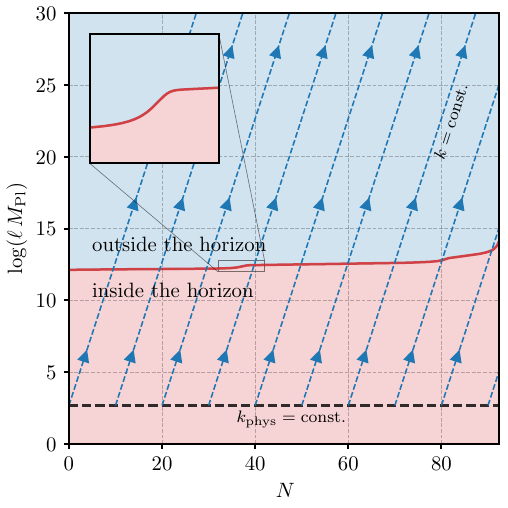}
\caption{\label{fig:inj_scheme} Mode injection scheme and evolution of the inflationary horizon (in the red solid curve). Time-translation invariance allows us to initialize each constant-$k$ mode (in light blue dashed lines) from a common initial surface of fixed physical wavelength (in a dashed black line). The inset in the upper-left corner illustrates variations in the horizon scale, which arise as a consequence of the sharp turns in the background trajectory shown in the right panel of \Figref{fig:def_geom_V_traj}.}
\end{figure}
}

\newcommand{\ctchecks}{%
\begin{figure*}[t!]
\centering
\includegraphics{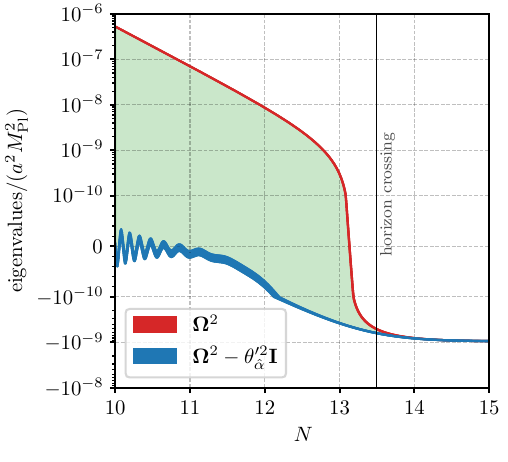}
\hfill
\includegraphics{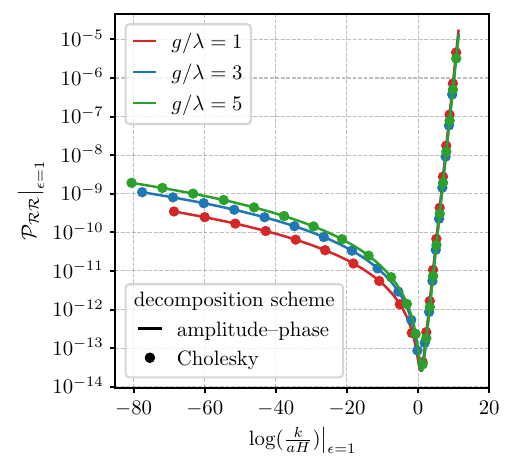}
\caption{\label{fig:freq_full_ps} Validation of the 
amplitude-phase separation method. Left panel: Evolution of the eigenvalues of the 
effective oscillation frequency before (in solid red) and after (in solid blue) applying the scale separation procedure. The eigenvalues 
are reduced by several orders of magnitude, allowing for larger evolution timesteps and significantly improving the 
computational efficiency. This supports the conclusion that the procedure described in section \ref{sec:method} effectively separates the largest oscillation scales from the system. 
Right panel: Power spectrum 
of primordial curvature fluctuations in an inflationary scenario 
sourced by the potential $V(\varphi^1,\varphi^2)=\frac{\lambda}{4}(\varphi^1)^4+\frac{g}{2}(\varphi^1)^2(\varphi^2)^2$. For sufficiently smooth potentials, the power spectrum shows that the results obtained using the scale-separation method agree with those from the Cholesky decomposition approach of \cite{GalvezGhersi:2016wbu} across all wavenumbers $k$.}
\end{figure*}
}

\newcommand{\modevol}{%
\begin{figure*}[t!]
\centering
\includegraphics{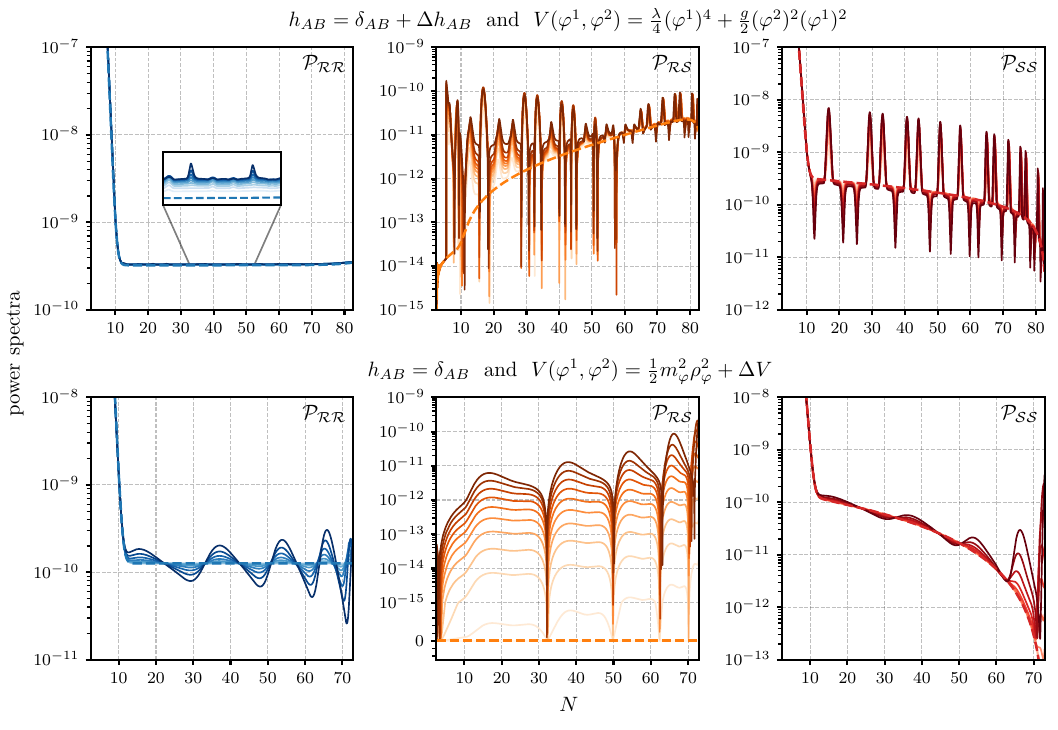}
\caption{\label{fig:mode_evol} Evolution of 
adiabatic, cross-correlation and isocurvature 
power spectra at fixed wavenumber in the presence of sharp turns in the background field evolution. The upper 
panels show the effects of progressively 
larger geometric deformations parameterized 
according to \eqref{eq:fld_met_ex} considering a smooth potential. The lower 
panels illustrate the deformations of the 
primordial spectra induced by increasingly 
large deviations in the potential, 
parameterized according to \eqref{eq:potential_st}. In both cases, darker colors correspond to larger departures from flat field-space geometry or from the quadratic potential, while dashed lines denote the undeformed cases where $\Delta V=0$ and $\Delta h_{AB}=0$. During mode evolution, these deviations can enhance the cross-correlation power to values comparable to those of the adiabatic and isocurvature modes. In contrast to the behavior observed in the Cholesky decomposition approach of \cite{GalvezGhersi:2016wbu} (see \Figref{fig:cholesky_problem}), the mode evolution here proceeds without interruption from dynamical instabilities.}    
\end{figure*}
}

\newcommand{\psk}{%
\begin{figure*}[t!]
\centering
\includegraphics{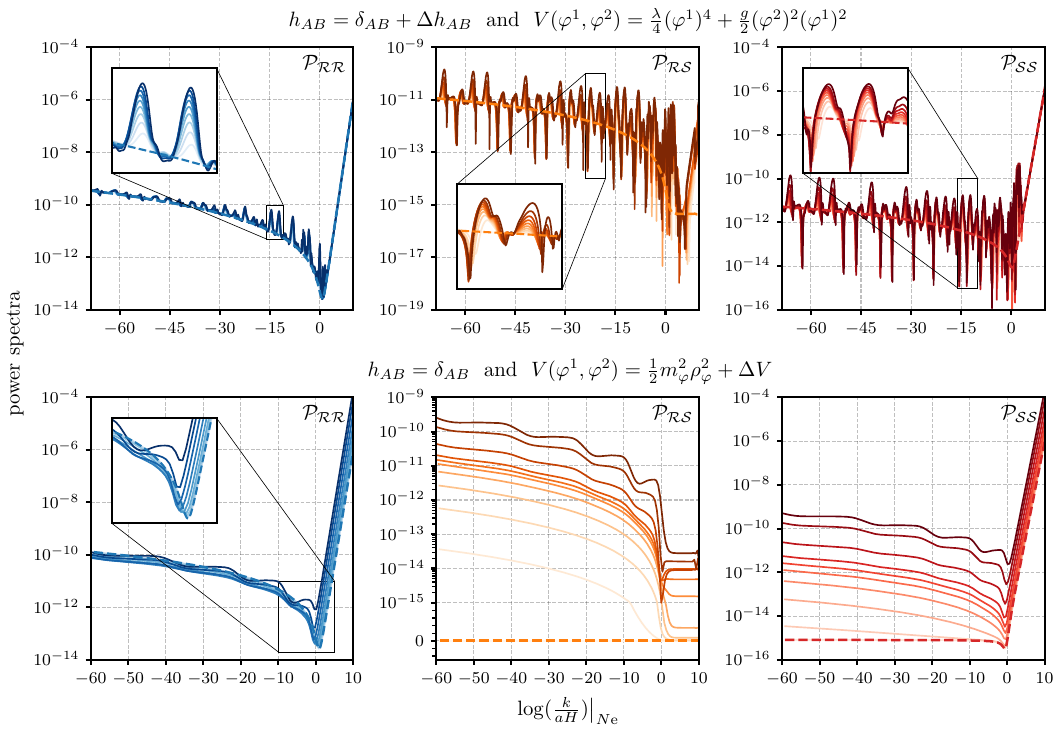}
\caption{\label{fig:ps_k} Adiabatic, cross-correlation, and isocurvature power 
spectra as functions of the comoving wavenumber $k$, 
corresponding to the geometric and potential deformations shown in 
\Figref{fig:mode_evol} and parameterized by Eqs.~\eqref{eq:fld_met_ex} and 
\eqref{eq:potential_st}, respectively. The upper panels display cases 
with nontrivial geometric deformations and a smooth quartic potential, while the lower 
panels show potential deformations with $\Delta h_{AB}=0$. In both cases, 
increasingly large deformations are represented by progressively darker 
curves. Depending on the field dependence of the deformation, the spectra may display either sharp or
smooth features across different $k$ bands.}    
\end{figure*}
}

\newcommand{\ellipses}{%
\begin{figure*}[t!]
\centering
\includegraphics{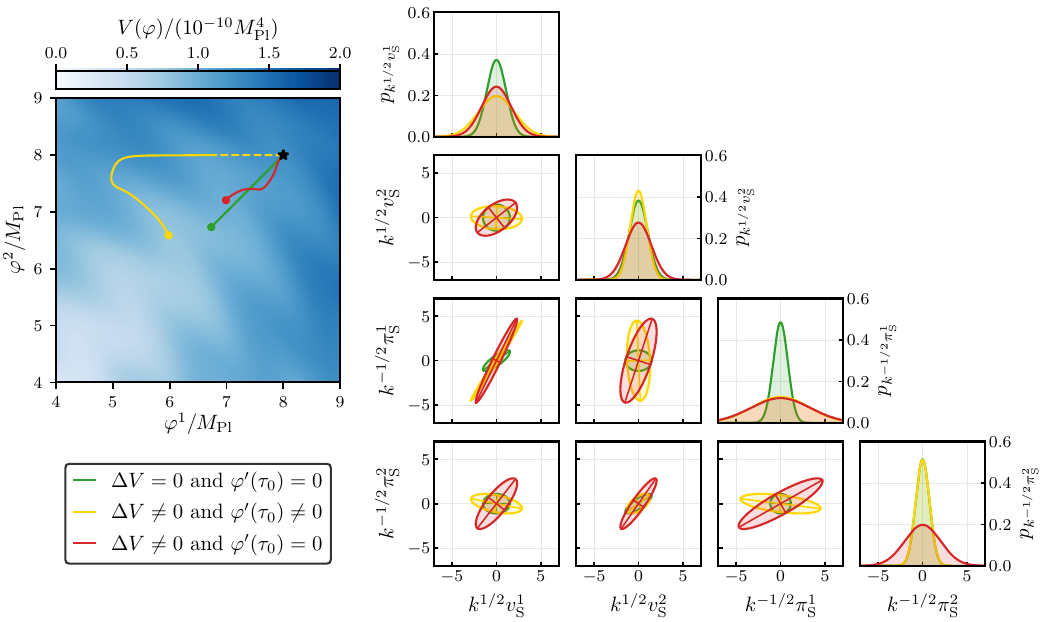}
\caption{\label{fig:ellipses} Evolution 
of the Wigner ellipses for different 
background field trajectories for the 
potential parameterization in 
\eqref{eq:potential_st}. Left panel: 
Background field evolution at a fixed instant of time in the 
$(\varphi^1,\varphi^2)$ plane for three 
cases: (a) $\Delta V=0$ with vanishing 
initial velocity (in green), (b) 
$\Delta V\neq 0$ with nontrivial initial 
field velocities (in yellow), and (c) 
$\Delta V\neq 0$ with vanishing initial 
velocities (in red). As in the right 
panel of \Figref{fig:def_geom_V_traj}, 
the presence of potential deformations 
bends the background trajectories, while 
nontrivial initial velocities can drive 
the fields toward potential walls, 
inducing sharper turns through 
reflections. Right panel: Triangular plot 
showing the projections of the Wigner 
ellipses corresponding to the three cases 
displayed in the left panel, evaluated at 
the same instant of time. In the diagonal subpanels, we 
show the marginalized Gaussian 
probability distribution functions $p$ 
associated with each dimensionless 
canonical variable. Due to the distinct 
coupling terms appearing in \eqref{eq:m_orig}, 
which dominate in each case, the ellipses 
deform during sharp turns and continue to 
evolve as the accelerated expansion 
proceeds. This manuscript includes two 
animations illustrating the effects of 
geometric and potential deformations.}    
\end{figure*}
}

\newcommand{\sixfieldsps}{%
\begin{figure*}[t!]
\centering
\includegraphics{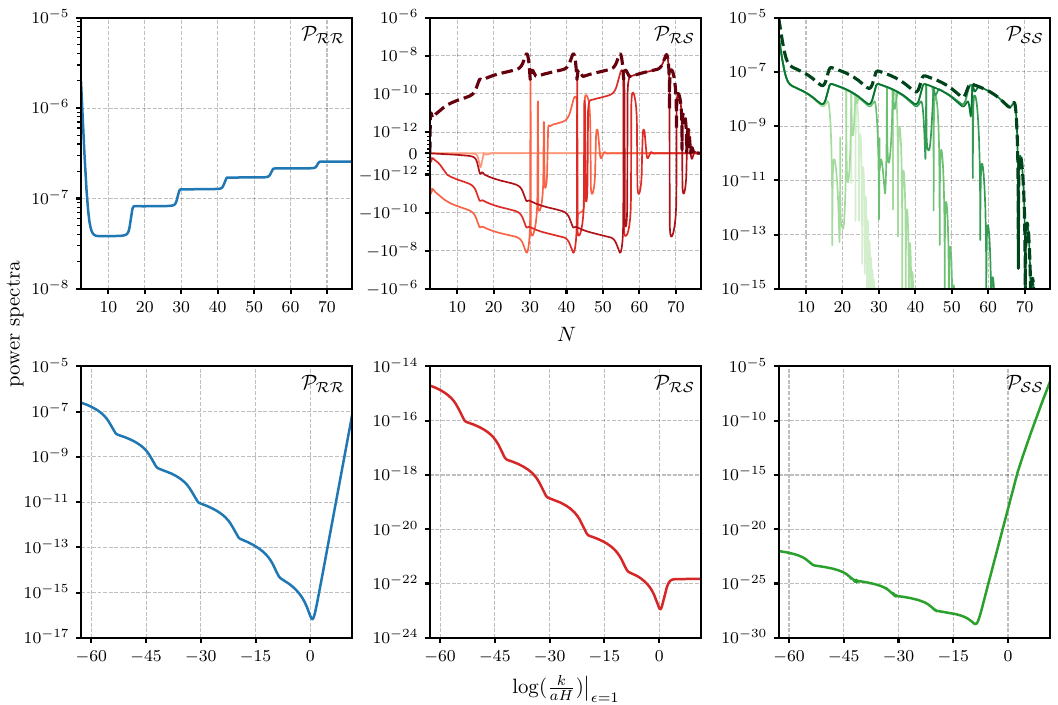}
\caption{\label{fig:6fields_ps} Adiabatic, cross-correlation, and isocurvature power spectra as functions of time and the dimensionless comoving wavenumber $k$, corresponding to the six-field 
elliptic potential in \eqref{eq:elliptic}. The upper panels display the time evolution of the 
spectra for a fixed $k$, consistent with the dynamics of a hierarchy of massive fields in which the
heaviest fields are the first to fall toward the origin. Contributions to the isocurvature and 
cross-correlation spectra from heavier fields are shown with progressively lighter colors. 
The lower panels show how the deformations in the mode evolution propagate to the $k$-dependent 
spectrum, producing bump-like features consistent with the sequential activation of 
lighter directions in the potential. In this specific case, the contribution from 
massive fields is so small that it cannot be represented in the spectra. None of the 
panels shows evidence of dynamical or numerical instabilities.}    
\end{figure*}
}

\newcommand{\renauxpetelmetric}{%
\begin{figure}[t!]
\centering
\includegraphics{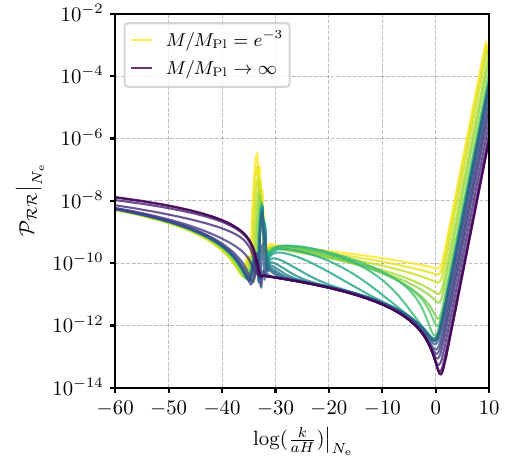}
\caption{\label{fig:renauxpetelmetric} Deformations 
of the primordial curvature power spectrum for the 
nontrivial field-space geometry in 
\eqref{eq:geom_dest_mt}, evaluated after 
$N_{\rm e}$ e-folds (\ie the end of 
inflation in the flat-metric case). $M$ controls the transition between 
potential- and geometry-dominated regimes. An ancillary 
animation shows the same effect evaluated at the end 
of inflation for each value of $M$, accounting for 
the corresponding shift in the total number of 
e-folds.}
\end{figure}
}

\begin{document}

\title{Optimized numerical evolution of perturbations across\\ sharp background trajectory turns in multifield inflation}

\author{Guillermo F.\ \surname{Quispe Pe\~na}\,\orcidlink{0009-0000-3775-6520}}
\email{gfq@sfu.ca}
\affiliation{Department of Physics, Simon Fraser University; 8888 University Dr. W, Burnaby, BC V5A 1S6}

\author{Johor D.\ \surname{Pe\~nalba Quispitupa}\,\orcidlink{0009-0004-2594-139X}}
\email{jpenalbaq@uni.pe}
\affiliation{Facultad de Ciencias, Universidad Nacional de Ingenier\'ia; Av. Tupac Amaru 210, Rímac, Lima, Per\'u}

\author{Jos\'e T.\ \surname{G\'alvez Ghersi}\,\orcidlink{0000-0001-7289-3846}}
\email{jgalvezg@utec.edu.pe}
\affiliation{Departamento de Ciencias, Universidad de Ingenier\'ia y Tecnolog\'ia; Jr. Medrano Silva 165, Barranco 15063, Lima, Per\'u}

\date{\today}

\hypersetup{pdfauthor={Quispe Pe\~na, Pe\~nalba Quispitupa and G\'alvez Ghersi}}

\begin{abstract}
Features in the primordial power spectrum 
require numerical methods that are both accurate 
and scalable across the wide class of multifield 
inflationary models that produce them. Sharp 
turns in the background trajectories, induced by 
either potential or geometric effects, render 
these computations particularly challenging. In 
this work, we introduce an efficient method for 
evolving primordial scalar fluctuations, 
requiring time steps comparable to those used for 
the background evolution. We demonstrate that the 
method accurately tracks perturbations through 
rapidly turning trajectories in arbitrary 
field-space geometries, enabling 
systematic exploration of spectral features 
across diverse multifield scenarios. Our 
approach scales robustly to large 
numbers of degrees of freedom, providing a 
reliable computational framework for probing 
regimes that significantly depart from slow-roll 
or slow-turn dynamics.
\end{abstract}
\maketitle

\section{Introduction}
\setcounter{equation}{0}

Inflation, at the perturbative level \cite{Guth:1980zm, Bardeen:1983qw, 
Mukhanov:1990me, Starobinsky:1992ts}, is a compelling mechanism that 
explains the origin of the initial conditions of the early universe 
through quantum-mechanical fluctuations. These initial conditions are 
consistent with the levels of inhomogeneity observed in the Cosmic 
Microwave Background (CMB) \cite{Smoot:1998jt, Bennett_2013, 
Planck:2018jri}. Forthcoming high-resolution CMB experiments 
\cite{SimonsObservatory:2018koc, CMB-S4:2020lpa, Belkner:2023duz, Hertig:2024adq} will provide tighter constraints on inflation and improve prospects for detecting primordial non-Gaussianity and gravitational-wave backgrounds using polarization data with reduced dust contamination.

Recent theoretical developments seek to embed inflation within high-energy 
frameworks such as supergravity and string theory 
\cite{Kallosh:2016gqp, Kallosh:2019jnl, Kallosh:2022ggf, Sarkar:2024jxz}. 
In many realizations, the inflationary dynamics involve multiple scalar 
fields evolving on a nontrivial field-space manifold. The resulting 
curvature modifies the effective oscillation frequencies and couplings of 
perturbations, thereby affecting the evolution of fluctuation modes through 
their coupling to the background trajectory. In particular, such geometric 
effects can lead to phenomena such as geometric destabilization 
\cite{Renaux-Petel:2015mga}. Nontrivial field dynamics may also arise 
from the structure of the inflationary potential itself 
\cite{Garcia-Bellido:2011kqb, Linde:2016uec, Aragam:2019omo, Bhattacharya:2022fze}. 
In such inflationary settings, suitable potential deformations can induce 
sharp turns in the trajectory followed by the background fields, 
generating localized features in the primordial curvature spectrum. In 
many-field models containing heavy fields transverse to the inflationary 
trajectory, these degrees of freedom may be integrated out to obtain an 
effective description of the adiabatic mode, whose regime of validity is 
controlled by the strength and duration of the turn 
\cite{Achucarro:2010da,Langlois:2008mn,Cespedes:2012hu}. Sharp turns may 
also leave characteristic imprints on the statistics of curvature 
perturbations, including specific forms of primordial non-Gaussianity 
\cite{Carlson:2025krq}. Both field-space geometry and potential-induced 
trajectory bending therefore provide natural mechanisms for departures 
from slow-roll dynamics and may leave observable imprints in the CMB and 
in large-scale structure formation. An additional motivation for studying 
scenarios with strongly curved trajectories is that they can produce a 
sufficient number of e-folds while keeping field excursions sub-Planckian 
\cite{Martin:2000xs}.

Perturbation modes injected from deep inside the horizon typically 
oscillate on very short timescales, and accurately resolving these 
oscillations can make their numerical evolution computationally expensive. 
To mitigate this cost, we originally proposed a method to separate the 
fast and slowly evolving components of the modes in both single-field 
\cite{GalvezGhersi:2018haa} and multifield \cite{GalvezGhersi:2016wbu} 
inflationary models. This approach relies on the observation that the 
rapidly oscillating phases are not required to evolve the covariance 
matrix of Gaussian fields, allowing the dynamics to be recast in terms of 
slowly varying quantities. However, a more careful review of this 
method reveals an important limitation: isolating the fast and slow 
oscillation scales within a mode-correlator transport scheme 
is not sufficient to solve the perturbation dynamics in situations where 
the background trajectories undergo arbitrarily fast changes in their 
evolution. In such scenarios, the strategy of \cite{GalvezGhersi:2016wbu} 
is not sufficient to faithfully resolve the full mode dynamics. 
Hence, the goal of this work is to develop and validate a framework for 
the evolution of multifield fluctuations undergoing nontrivial background 
evolution,  particularly in regimes where standard approaches face difficulties. 
Accordingly, we focus on demonstrating the robustness of the method 
across a range of scenarios, rather than providing a detailed physical 
interpretation of the resulting spectral features.

To study the evolution of perturbations, we first perform 
coordinate transformations in field space that recast the mode equations 
into a form analogous to the Mukhanov-Sasaki equation for single-field 
inflation \cite{Baumann:2009ds}. This coordinate choice is reminiscent of 
a parallel-transport gauge in General Relativity, which is used to map 
quantities between accelerated and inertial frames. In this particular 
frame, the symplectic structure of the mode functions remains consistent 
with the canonical commutation relations and allows us to rewrite the mode equations of motion in terms of phases and mode amplitudes. The key observation is that the amplitude equations of motion can be 
expressed as a nonlinear second-order system of ordinary differential equations 
whose effective oscillation frequency is suppressed by several orders of magnitude. This 
suppression enables the use of larger time steps, thereby improving the 
efficiency of numerical evolution schemes. 

With the dynamical system at 
hand, we demonstrate that the evolution of perturbations is stable in 
scenarios where the dynamics of the background fields is modified by 
(a) field-space curvature, (b) potential deformations inducing sharp 
turns in the background trajectories, or both. Such scenarios may generate 
features in the primordial spectrum of curvature fluctuations, and exhibit
nontrivial cross-correlations with the isocurvature modes. This approach is formulated for an arbitrary number of degrees of freedom, including those in which sharp turn effects play a prominent role.

The remainder of this paper is organized as follows. In 
\secref{sec:method}, we introduce our fast and slow decomposition 
through an amplitude-phase parametrization of the mode functions. We 
begin by reviewing the foundations of inflationary 
dynamics at second-order in perturbation theory and by rewriting the 
perturbation mode equations of motion in the parallel-transport gauge in 
field space. In these coordinates, we discuss 
the symplectic structure of the dynamical system, and use it to construct 
our dynamical amplitude-phase decomposition scheme. We then describe the 
relevant cosmological observables and build the initial conditions that 
complete the dynamical system. In \secref{sec:results}, we present the 
results of a numerical implementation of this approach, focusing on 
scenarios in which sharp turns of the background trajectories are sourced 
by field-space curvature effects (including the case of geometrical 
instabilities presented in \cite{Renaux-Petel:2015mga}), by 
various structures in the field potential, or 
by an arbitrary combination of the two effects. Finally, in 
\secref{sec:disc} we discuss and conclude.

\section{Dynamical setup}
\label{sec:method}
\setcounter{equation}{0}

In this section, we introduce our dynamical fast-slow decomposition scheme. To do so, we first establish the conventions and notations required to define the inflationary setup. In particular, we present the equations of motion governing the background field evolution and specify the field-space and spacetime geometries on which the perturbations propagate. We consider a model with $N_{\rm f}$ real scalar fields in $(3+1)$-dimensional spacetime and adopt the metric signature $(-,+,+,+)$. Throughout this manuscript, we work in natural units with $\hbar = c = 1$, and express all the energy and inverse length scales in terms of the reduced Planck mass, $M_{\mathrm{Pl}} \equiv (8\pi G)^{-1/2}$. Spacetime indices are denoted by lowercase Greek letters, and field-space indices by uppercase Latin letters.

The dynamics are governed by an action describing the dynamics of $N_{\rm f}$ scalar fields minimally coupled to gravity, defined on a nontrivial field-space geometry that endows the fields with a noncanonical kinetic term
\begin{align}\label{eq:EH_action}
    S = \int \mathrm{d}^{4}x \sqrt{-g} \left( \frac{M^2_{\mathrm{Pl}}}{2} R - \frac{1}{2}  h_{AB} \partial_{\mu}\phi^{A} \partial^{\mu}\phi^{B} - V \right) \,,
\end{align}
where $g \equiv \mathrm{det}(g_{\mu\nu})$, $R$ is the Ricci scalar constructed from the spacetime metric $g_{\mu\nu}$, and $V(\phi)$ denotes the scalar field potential. Here and throughout, uppercase Latin indices  label coordinates in the $N_{\rm f}$-dimensional field space. As stated before, this field space is endowed with a metric $h_{AB}(\phi)$, which determines the kinetic structure of the theory. Indices are raised and lowered with $h^{AB}$ and $h_{AB}$, respectively. Throughout this manuscript, the norm of a field-space vector $X^{A}$, induced by the metric $h_{AB}$, is denoted by $|X| = \sqrt{h_{AB} X^{A} X^{B}}$.

To investigate the evolution of perturbations during inflation, we expand around a spatially flat Friedmann-Lemaître-Robertson-Walker (FLRW) background. The unperturbed metric is characterized by the scale factor $a(t)$, and scalar perturbations are introduced in the 
spacetime metric as
\begin{equation}
    \mathrm{d}s^{2} = -(1+2A) \, \mathrm{d}t^{2} + 2a^{2} \partial_{i} B \, \mathrm{d}x^{i} \, \mathrm{d}t + a^{2} \delta_{ij} \, \mathrm{d}x^{i} \, \mathrm{d}x^{j} \, ,
\end{equation}
where $A(\mathbf{x},t)$ and $B(\mathbf{x},t)$ are the scalar perturbation variables in the 
spatially-flat gauge. This choice of gauge eliminates perturbations in the purely spatial part 
of the metric, simplifying the description of field fluctuations.

In the scalar sector, each field component 
$\phi^{A}$ is decomposed into a homogeneous 
background trajectory and small inhomogeneous 
fluctuations:
\begin{equation}
\label{eq:nv_pert}
    \phi^{A}(\mathbf{x},t) \equiv \varphi^{A} (t) + \delta\phi^{A}(\mathbf{x},t) \, ,
\end{equation}
where this standard first-order perturbative expansion provides the organizing framework for the remainder of the section. In \subsecref{subsec:back}, we show that the evolution of the background fields $\varphi^A(t)$ defines a trajectory on the curved field-space geometry. In \subsecref{subsec:eq_motion}, we introduce the equations of motion governing the perturbations 
$\delta\phi^A$, which characterize the linear response of the system to deviations that are either parallel or orthogonal to this background trajectory. In that subsection, we further demonstrate explicitly how the curvature of the field-space geometry induces additional terms through which the background evolution sources and drives the perturbation dynamics. Finally, in the subsequent subsections \ref{subsec:sym_struc_mode_func}--\ref{subsec:init_cond_schm}, we present the decomposition scheme that isolates fast and slow degrees of freedom and efficiently resolves the evolution of the fluctuation modes.

\trajback

\subsection{Background equations}\label{subsec:back}

The main objective of this manuscript is to study the dynamics of perturbations as the background trajectories undergo abrupt changes. With this in mind, in this subsection we illustrate the dynamics of background fields as these undergo deformations caused by nontrivial geometries and by potential deformations. To do so, we first consider that, at the background level, it is reasonable to
assume spatial homogeneity and consider $N_{\rm f}$ scalar fields $\varphi^{A}(t)$ taking values in an $N_{\rm f}$-dimensional field manifold. The field-space metric $h_{AB}$ determines the kinetic inner product and may depend on the fields. Dots denote derivatives with respect to cosmic time $t$ and the Hubble rate is $H \equiv \dot{a} / a$.

The covariant equation of motion for the homogeneous fields is the covariant generalization of a damped nonlinear harmonic oscillator,
\begin{equation}
    \mathcal{D}_{t}\dot{\varphi}^{A} + 3H \dot{\varphi}^{A} + h^{AB} \partial_{B} V = 0 \, ,
    \label{eq:back_field}
\end{equation}
where $\partial_{A}$ denotes partial differentiation with respect to the $A$-th field component and indices are raised with the inverse metric $h^{AB}$. The operator $\mathcal{D}_{t}$ represents the covariant derivative projected along the background field velocity, and acts on a field-space vector $X^{A}(t)$ as  $\mathcal{D}_{t} X^{A} = \dot{X}^{A} + \Gamma^{A}_{BC} \dot{\varphi}^{C} X^{B}$, with $\Gamma^{A}_{BC}$ the Christoffel symbols of the Levi-Civita connection on the field-space manifold, $\Gamma^{A}_{BC} = \frac{1}{2} h^{AD} ( \partial_{C} h_{DB} + \partial_{B} h_{DC} - \partial_{D} h_{BC} )$. In particular, $\mathcal{D}_{t} \dot{\varphi}^{A} = \ddot{\varphi}^{A} + \Gamma^{A}_{BC} \dot{\varphi}^{B} \dot{\varphi}^{C}$ is the covariant acceleration of the background trajectory in field space. With respect to 
the evolving background fields, the Friedmann constraint equations then read as
\begin{subequations}
    \begin{align}
        H^{2} &= \frac{1}{3 M_{\mathrm{Pl}}^{2}} \! \left( \frac{1}{2} h_{AB} \dot{\varphi}^{A} \dot{\varphi}^{B} + V \right) 
         \label{eq:hubble}\, , \\
        \dot{H} &= - \frac{1}{2 M_{\mathrm{Pl}}^{2}} h_{AB} \dot{\varphi}^{A} \dot{\varphi}^{B} \, .
        \label{eq:h_dot}
    \end{align}
\end{subequations}
These two equations fully determine the homogeneous FLRW background once the fields $\varphi^{A}(t)$ and their velocities are specified.

It is essential to emphasize the role of the field-space geometry in the first-order perturbative treatment of multifield inflation. When the field-space metric $h_{AB}(\varphi)$ is non-trivial, the covariant acceleration term proportional to $\Gamma^{A}_{BC}$ in the background field equations encodes geometric “forces” that bend the background trajectory in field space. Field-space geometry also influences the rate at which the Hubble parameter evolves. At the level of perturbations, field-space curvature introduces additional couplings between isocurvature (entropy) and adiabatic fluctuation modes, which may lead to instabilities, as discussed in \cite{Renaux-Petel:2015mga}. 

Field-space geometry and potential deformations can induce sharp turns in the background trajectories. In line with the objectives of this paper, which are to explore the effects of such sharp turns on the evolution of field fluctuations, we illustrate the evolution of representative background field trajectories in \Figref{fig:def_geom_V_traj} for inflationary models with two fields. With the equations of motion at hand, in the left panel, we evolve  the background fields in the presence of a smooth nonlinear potential. Despite the smoothness of the potential, the trajectories exhibit abrupt changes in their evolution, sourced by the nontrivial diagonal field-space metric
\begin{equation}\label{eq:fld_met_ex}
    h_{AB} = \delta_{AB} + \Delta h_{AB}
\end{equation}
with
\begin{equation}
\Delta h_{AB} \equiv \delta^{2}_{A} \delta^{2}_{B} \sum_{i=1}^{N_{\rm b}} A_{i} \exp \bigg[ - \frac{1}{2} \frac{(\varphi^{1}-x_{i})^{2}}{\sigma^{2}} \bigg] \, .
\end{equation}
Here the dependence of the metric component $h_{22}$ on the 
background fields enables significant departures from the standard 
background evolution (in red) at different instants of inflation, 
leading to a variety of deformations along the isocurvature or adiabatic directions. It is clear that the field metric in \eqref{eq:fld_met_ex} deviates from a Euclidean form due to the presence of a series of $N_{\rm b}$ narrow  Gaussian bumps. To produce the trajectories in the left panel of \Figref{fig:def_geom_V_traj}, we considered $N_{\rm b}=30$ bumps of fixed width $\sigma=0.06 M_{\rm Pl}$ and located along the $\varphi^1$ axis in $x_i/M_{\rm Pl}\in[-19 , 19]$. The amplitude of each Gaussian bump is chosen to be either positive $(A_i=40)$ or negative $(A_i=-0.75)$, producing peaks and troughs in the resulting background trajectories. In the following sections, we illustrate how suitably chosen field-space geometries can accommodate different types of features in the primordial power spectrum.

In addition to corrections from the field-space geometry, the scalar potential can also induce sharp turns in the field trajectories \cite{Bartolo:2013exa,Bhattacharya:2022fze}. In the right panel of \Figref{fig:def_geom_V_traj}, we show the evolution of a background trajectory embedded in the two-field potential
\begin{equation}\label{eq:potential_st}
    V(\varphi^{1},\varphi^{2}) = \frac{1}{2} m_{\varphi}^{2}\rho_{\varphi}^{2} + \Delta V \, , 
\end{equation}
where
\begin{equation}
    \Delta V \equiv \lambda_{\rm V}^{4}  \cos^2 \bigg[  \frac{f_{1}}{2} \upsilon_{\varphi} - \frac{f_{2}}{2} \sin\Big( f_{3} \rho_{\varphi} \Big) \bigg] 
\end{equation}
represents a deformation of the otherwise quadratic potential. Here $\rho^2_{\varphi}\equiv (\varphi^{1})^{2}+(\varphi^{2})^{2}$ denotes the radial field coordinate, while $\upsilon_{\varphi}\equiv\tan^{-1}(\varphi^{2}/{\varphi^{1})}$ is the angular variable in field space. For the example shown in the figure, the parameters are chosen as $m_{\varphi}=1.4\times 10^{-6}M_{\rm Pl}$, $\lambda_{\rm V}=2\times 10^{-3}M_{\rm Pl}$, $f_1=4$, $f_2=2$ and $f_3=M^{-1}_{\rm Pl}$. 
The purpose of this potential is to provide an illustrative example in which background trajectories exhibit pronounced bending at multiple stages of their evolution (the curve in dashed black lines) and are sensitive to deformations by larger field velocities (in solid red). This is relevant when exploring the effects of field trajectories which are distant from the attractor at early times. 

Another motivation for considering sharp turns in the potential is that they allow for inflationary models in which sufficient e-folds can be generated without large field excursions, which could be a possible solution for the trans-Planckian problem \cite{Martin:2000xs}. Although not shown here, we find that both geometric and potential deformations can be used to construct such models. In future work, we will explore these scenarios in greater detail and compute the associated field fluctuations in an efficient and numerically stable manner.

\srparams

While the trajectories in \Figref{fig:def_geom_V_traj} illustrate qualitatively how geometric and potential deformations can bend the background trajectory, a quantitative characterization of their evolution requires parameters describing both changes in the field-space speed and changes in the direction of motion. The first slow-roll parameter,
\begin{equation}
    \epsilon \equiv - \frac{\dot{H}}{H^{2}} = \frac{|\dot{\varphi}|^{2}}{2 M_{\rm Pl}^{2} H^{2}} \, ,
    \label{eq:epsilon_def}
\end{equation}
quantifies the departure from a perfect de Sitter expansion. By itself, however, $\epsilon$ carries no information about the direction along which the fields accelerate in field space.

To capture this directional information, we introduce an orthonormal field-space vielbein $e_{\cal X}^{A}$ adapted to the background trajectory, with label $\mathcal{X} \in \{\mathcal{R}, \mathcal{S}_{1}, \dots, \mathcal{S}_{N_{\rm f}-1} \}$, satisfying $h_{AB} \, e_{\cal X}^{A} e_{\cal Y}^{B} = \delta_{\cal XY}$. The adiabatic direction $e_{\cal R}^{A} \equiv \dot{\varphi}^{A} / |\dot{\varphi}|$ is tangent to the background trajectory, while the entropic directions $e_{\mathcal{S}_{a}}^{A}$ span its orthogonal complement in field space; this same basis is used in \subsecref{subsec:two_pt_pow_spec} to project the field perturbations onto curvature and isocurvature components. Projecting the covariant acceleration $\mathcal{D}_{t}\dot{\varphi}^{A}$ onto this basis defines the two additional kinematic parameters commonly used in this context,
\begin{subequations}
    \begin{align}
        \eta_{\parallel} &\equiv \frac{h_{AB} \, e_{\cal R}^{A} \, \mathcal{D}_{t}\dot{\varphi}^{B}}{H |\dot{\varphi}|} \, , \\
        \eta_{\perp} &\equiv \sqrt{ \sum_{a=1}^{N_{\rm f} - 1} \bigg( \frac{h_{AB} \, e_{\mathcal{S}_{a}}^{A} \, \mathcal{D}_{t}\dot{\varphi}^{B}}{H |\dot{\varphi}|} \bigg)^{2} } \, ,
    \label{eq:eta_par_eta_perp}
    \end{align}
\end{subequations}
where $\eta_{\parallel}$ measures the fractional acceleration (or deceleration, according to its sign) of the fields along the direction of motion, and $\eta_{\perp}$ is the turn rate, i.e.\ the rate at which the trajectory bends away from a field-space geodesic, both measured in units of $H$. Because the $e_{\cal X}^{A}$ are orthonormal, these definitions satisfy the exact identity
\begin{equation}
    |\mathcal{D}_{t}\dot{\varphi}|^{2} = H^{2} |\dot{\varphi}|^{2} \left( \eta_{\parallel}^{2} + \eta_{\perp}^{2} \right) \, ,
\end{equation}
so that $\eta_{\parallel}$ and $\eta_{\perp}$ are the components of the covariant acceleration parallel and orthogonal to the trajectory, respectively, measured in units of Hubble time. 

We distinguish the slow-roll and slow-turn conditions throughout this work. We refer to a background as being in the slow-roll regime when $\epsilon \ll 1$ and $|\eta_{\parallel}| \ll 1$, so that the expansion is close to de Sitter and both the field-space speed and $\epsilon$ vary slowly over a Hubble time. Independently, the trajectory is in the slow-turn regime when $\eta_{\perp} \ll 1$, for which the adiabatic direction changes only slightly during one e-fold. The combined slow-roll and slow-turn regime therefore requires all three quantities $\epsilon$, $|\eta_{\parallel}|$, and $\eta_{\perp}$ to be much smaller than unity. This distinction is important because a background may remain inflating, and may even satisfy the slow-roll conditions, while violating the slow-turn condition, either transiently or throughout a sustained rapid-turn phase \cite{Wolters_2024,Anguelova_2023}.

We illustrate this behavior with the two-field quadratic potential
\begin{equation}
    V(\varphi^{1},\varphi^{2}) = \frac{1}{2} m_{1}^{2} (\varphi^{1})^{2} + \frac{1}{2} m_{2}^{2} (\varphi^{2})^{2} \, ,
    \label{eq:elliptic-potential}
\end{equation}
on a Euclidean field-space metric $h_{AB} = \delta_{AB}$, for which the mass ratio $m_{2}/m_{1}$ controls the eccentricity of the potential and, in turn, the sharpness with which the trajectory bends as it relaxes from generic initial conditions onto the valley aligned with the lighter field. \Figref{fig:slow_roll_params} shows the evolution of $\epsilon$, $|\eta_{\parallel}|$, and $\eta_{\perp}$ obtained by numerically integrating \eqref{eq:back_field} for different mass ratios $m_{2}/m_{1}$. For $m_{2}/m_{1} = 1$ the potential has azimuthal symmetry, the trajectory is a straight line in field space, and $\eta_{\perp} \equiv 0$ identically, while $\epsilon$ and $\eta_{\parallel}$ evolve smoothly, without any localized feature. As the mass ratio is increased, the heavier direction relaxes on a shorter timescale than the lighter one, forcing the trajectory to bend sharply toward the shallow valley; correspondingly, $\epsilon$ develops a positive peak that becomes progressively narrower and more pronounced as $m_{2}/m_{1}$ increases, while $\eta_{\parallel}$ and $\eta_{\perp}$ exhibit increasingly large and localized spikes coincident with the peak in $\epsilon$, with $\eta_{\parallel}$ changing sign across the turn. This provides, in a minimal and controlled setting, a realization of the notion of a sharp turn introduced above: a localized transient during which $\eta_{\perp}$ becomes sufficiently large to violate the slow-turn condition $\eta_{\perp}\ll1$ that otherwise holds along the trajectory. Sharp turns of this type can transmit the effects of heavy isocurvature modes to the curvature perturbation and generate pronounced features in the primordial power spectrum \cite{Cespedes:2012hu}. To evaluate the robustness of the method for evolving field perturbations proposed in this paper, however, we also consider other scenarios in which sharp-turn events occur repeatedly throughout the evolution of the background trajectory.

\CDfail

\subsection{Perturbation equations and the parallel-transport gauge}\label{subsec:eq_motion}

The strategy of resolving the evolution of 
perturbations in multifield inflationary 
models by separating fast and slow degrees of 
freedom (an approach that underlies the main 
objective of this work) is not new. In 
\cite{GalvezGhersi:2016wbu}, we introduced a 
separation method based on solving the 
transport equations for the two-point 
correlation functions in Fourier space. A 
careful reevaluation of those results shows that 
the evolution of field correlators presented in that paper becomes unstable when the background trajectories undergo sufficiently abrupt changes during their evolution. 
To illustrate this, consider the elliptic potential of \eqref{eq:elliptic-potential}, where, as discussed above, the eccentricity of the potential (parameterized by the mass ratio $m_2/m_1$) controls the sharpness of the turn in the background trajectories. 
\Figref{fig:cholesky_problem} illustrates how the amplitude–phase decomposition 
introduced in this work successfully resolves the mode dynamics by evolving the 
correlators in terms of variables that do not require explicitly tracking rapidly 
oscillating phases (dashed lines). By contrast, the approach of 
\cite{GalvezGhersi:2016wbu}, based on the evolution of the Cholesky factors of 
the covariance matrix (solid lines), becomes unstable in regimes 
of rapid background evolution, leading to an 
interruption of the solid lines. For sufficiently 
large mass ratios, this prevents the evolution from 
reaching the end-of-inflation surface, and hence 
the primordial spectrum cannot be fully resolved.
We have checked that the instabilities of the 
Cholesky decomposition approach are triggered by the equations of motion, and are not an 
artifact of a specific numerical implementation. 
As the mass ratio decreases, corresponding to a milder turn, this interruption 
occurs at later times, which implies that the Cholesky decomposition approach is 
only valid for sufficiently smooth potentials. The inset in the upper-left corner
of the figure highlights a nontrivial mode evolution that is distinct from the 
mode freezing characteristic of single-field inflation. This post-turn evolution 
emerges immediately after the trajectory bends for larger values of the mass 
ratio and in the context of other inflationary models, it may imprint features in the adiabatic, isocurvature, and 
cross-correlation power spectra. 

In an analogous manner to potential-induced features, nontrivial field-space 
geometry can generate sharp turns in the background trajectory even when the 
potential is otherwise smooth. This suggests that the limitations of the approach 
of \cite{GalvezGhersi:2016wbu}, identified above, may also arise in geometrically 
induced turning regimes. Motivated by this observation, we develop an amplitude–phase 
decomposition framework that remains operational and numerically stable in the presence 
of such nontrivial geometries, enabling a controlled evolution of field perturbations 
through sharp turns without introducing dynamical instabilities.

It is useful to distinguish the present method from existing correlator-transport frameworks, including \textsc{PyTransport} \cite{Mulryne:2016mzv} and the cosmological-flow formalism \cite{Pinol:2023oux}. These approaches evolve equal-time correlation functions directly, controlling the computational cost of subhorizon oscillations through a hybrid strategy: analytic solutions from the adiabatic expansion are used to fix the correlators at a time sufficiently deep inside the horizon, from which they are evolved numerically through horizon crossing. In addition, the antisymmetric part of the correlator is fixed by the canonical commutation relations and need not be evolved; the real symmetric part, however, may still contain rapid oscillations generated by nonadiabatic mode mixing. By contrast, the present decomposition is performed at the level of the mode functions themselves: the dominant phase is constrained by the conserved symplectic structure, while only the amplitude variables are evolved. Its main advantage is therefore expected in problems involving long subhorizon evolution, large frequency hierarchies, or sharp background turns — precisely the regime that is the focus of this manuscript.

With the motivation properly introduced, we develop in this subsection our separation framework 
at first order in perturbation theory considering a nontrivial field-space geometry. In this 
context, it is essential to note that when the scalar fields $\phi^{A}$ span a curved 
field-space manifold, their perturbations must be defined in a covariant manner. The naive 
coordinate difference $\delta\phi^{A}(\mathbf{x},t) = \phi^{A}(\mathbf{x},t) - \varphi^{A}(t)$ from equation 
\eqref{eq:nv_pert} does not transform as a vector under field redefinitions, and therefore, it 
is not an appropriate perturbation variable when the field-space metric $h_{AB}$ is nontrivial. 
To obtain a quantity that transforms covariantly, one defines the perturbation 
$Q^{A}(\mathbf{x},t)$ through the exponential map on field space. This map transports the background 
field $\varphi^{A}(t)$ to the full field configuration $\phi^{A}(\mathbf{x},t)$,
\begin{equation}
    \phi^{A}(\mathbf{x},t) = [\mathrm{exp}_{\varphi(t)} Q(\mathbf{x},t) ]^{A} \, ,
\end{equation}
which implies that $Q^{A}(\mathbf{x},t)$ is the generator of displacements at $\varphi^{A}(t)$ whose 
geodesic, generated by the field-space connection $\Gamma^{A}_{BC}$, connects the background 
field with its target value $\phi^{A}(\mathbf{x},t)$.

Consistent with the approach in \cite{Jinn-Ouk_Gong_2011}, we consider a truncation of the exponential map, which yields the relation between the covariant and coordinate perturbations:
\begin{equation}
    \delta\phi^{A} = Q^{A} - \frac{1}{2} \Gamma^{A}_{BC} Q^{B} Q^{C} + [\mathcal{O}(Q^3)]^A \, ,
\end{equation}
so that at linear order one may identify $Q^{A} \simeq \delta\phi^{A}$, while the nonlinear completion ensures covariance of the full-perturbative expansion. With this definition at hand, the action for perturbations (up to second order in $Q$) takes the invariant form
\begin{widetext}
\begin{equation}\label{eq:action-perturbations}
    S_{(2)} = \frac{1}{2} \int \mathrm{d}^{4}x \, a^{3} \bigg( h_{AB} \mathcal{D}_{t} Q^{A} \mathcal{D}_{t} Q^{B} - \frac{1}{a^{2}} h_{AB} \nabla Q^{A} \cdot \nabla Q^{B} - \mathcal{M}^{2}_{AB} Q^{A} Q^{B} \bigg) \, ,
\end{equation}
where $\nabla$ denotes the flat spatial gradient on the spatial hypersurfaces, and the dot denotes the standard Euclidean inner product. The effective mass-squared matrix governing the coupled fluctuations is given by
\begin{equation}
    \mathcal{M}^{2}_{AB} \equiv \mathcal{D}_{B} \mathcal{D}_{A} V - R_{ACDB} \dot{\varphi}^{C} \dot{\varphi}^{D} - \frac{1}{a^{3}M_{\mathrm{Pl}}^{2}} \mathcal{D}_{t} \! \left( \frac{a^{3}}{H} \dot{\varphi}_{A} \dot{\varphi}_{B} \right) \, , 
    \label{eq:m_orig}
\end{equation}
where $R_{ABCD}$ is the Riemann tensor of the field-space metric $h_{AB}$. The first term represents the Hessian of the potential projected covariantly on field space, the second term encodes curvature effects due to the geometry of field space, and the third term is a consequence of the coupling between perturbations and the gravitational background.

To analyze the perturbations more systematically, it is convenient to decompose them in Fourier modes, which are defined on spatially flat hypersurfaces. Using the convention $Q^{A} (\mathbf{x},t) = (2\pi)^{-3/2} \int \mathrm{d}^{3} \mathbf{k} \, \Phi^{A} (\mathbf{k},t) e^{i \mathbf{k} \cdot \mathbf{x}}$, the quadratic action \eqref{eq:action-perturbations} can be expressed in Fourier domain as
\begin{equation}\label{eq:ac_orig}
    S_{(2)} = \frac{1}{2} \int \mathrm{d}t \, \mathrm{d}^{3} \mathbf{k} \, a^{3} \bigg[  h_{AB} \mathcal{D}_{t} \Phi^{A} \mathcal{D}_{t} \bar{\Phi}^{B}  - \bigg( \frac{k^{2}}{a^{2}} h_{AB} + \mathcal{M}^{2}_{AB} \bigg) \Phi^{A} \bar{\Phi}^{B} \bigg] \, ,
\end{equation}
where bars denote complex conjugation, since $Q^{A}$ is real. The Euler-Lagrange equations for the Fourier amplitudes follow by varying this action with respect to $\bar{\Phi}^{A}$:
\begin{equation}
    \mathcal{D}_{t}^{2} \Phi^{A} + 3H\mathcal{D}_{t}\Phi^{A} +  \bigg( \frac{k^{2}}{a^{2}} \delta^A_{C} + h^{AB}\mathcal{M}^{2}_{BC} \bigg) \Phi^{C} = 0\,.
\end{equation}
For completeness, one can expand the covariant time derivatives in terms of the Christoffel symbols $\Gamma^{A}_{BC}$ associated with the field-space metric $h_{AB}$. In explicit component form, the mode equation becomes
\begin{equation}\label{eq:eq_init}
    \ddot{\Phi}^{A} + (2 \Gamma^{A}_{BC} \dot{\varphi}^{C} + 3H \delta^{A}_{B}) \dot{\Phi}^{B} + \left[ \frac{\mathrm{d}}{\mathrm{d}t} ( \Gamma^{A}_{BC} \dot{\varphi}^{C} ) + 3H \Gamma^{A}_{BC} \dot{\varphi}^{C} + \Gamma^{A}_{CD} \Gamma^{C}_{BE} \dot{\varphi}^{D} \dot{\varphi}^{E} + h^{AC} \! \left( \frac{k^{2}}{a^{2}} h_{CB} + \mathcal{M}^{2}_{CB} \right) \right] \! \Phi^{B} = 0 \, .
\end{equation}
\end{widetext}

To simplify the dynamics of perturbations and bring the action into a form amenable to 
canonical quantization, it is convenient to move from the coordinate basis $\{ \partial_{A} \}$ on field space to a local orthonormal frame $\{ e_{i} \}$. The orthonormal basis is defined by $e_{i} = \Lambda^{A}_{i} \partial_{A}$ where $\Lambda^{A}_{i}$ is a vielbein on the field-space manifold. The vielbein is required to be parallel-transported along the background trajectory $\varphi^{A}$, namely,
\begin{equation}
    \mathcal{D}_{t}\Lambda^{A}_{i} = \dot{\Lambda}^{A}_{i} + \Gamma^{A}_{BC} \dot{\varphi}^{C} \Lambda^{B}_{i} = 0 \,,
    \label{eq:transport_vielbein}
\end{equation}
so that the basis vectors flow along the motion of the background. By construction, the vielbein relates the curved field-space metric $h_{AB}$ to the flat Euclidean metric $\delta_{ij}$ through
\begin{equation}\label{eq:old-to-new-metric}
    \delta_{ij} = \Lambda^{A}_{i} \Lambda^{B}_{j} h_{AB} \, ,
\end{equation}
or equivalently,
\begin{equation}\label{eq:new-to-old-metric}
    h_{AB} = \Lambda^{i}_{A} \Lambda^{j}_{B} \delta_{ij} \, ,
\end{equation}
with $\Lambda_{A}^{i}$ denoting the inverse vielbein, satisfying $\Lambda^{A}_{i} \Lambda^{i}_{B} = \delta^{A}_{B}$ and $\Lambda^{i}_{A} \Lambda^{A}_{j} = \delta^{i}_{j}$. Throughout this manuscript, we use uppercase Latin indices for coordinate-basis components and lowercase Latin indices for orthonormal-frame components. It is straightforward to prove that $\delta^{ij}$ and $\delta_{ij}$ can be used to raise and lower indices in the orthonormal frame.

We introduce the canonically rescaled fields $V^{i}(\mathbf{x},\tau) \equiv a(\tau) \Lambda^{i}_{A}(\tau) Q^{A}(\mathbf{x},\tau)$ where $\tau$ follows the standard definition of conformal time $\mathrm{d}\tau = \mathrm{d}t / a$. The associated canonical momentum density is $\Pi_{i}(\mathbf{x},\tau) \equiv \delta S_{(2)} / \delta V^{\prime i} = \delta_{ij} V^{\prime j}(\mathbf{x},\tau)$, where primes denote derivatives with respect to conformal time. We further define the orthonormal-frame mass-squared matrix as the projection $\mathcal{M}^{2}_{ij} \equiv \Lambda^{A}_{i}\Lambda^{B}_{j} \mathcal{M}^{2}_{AB}$. With these conventions, the quadratic action and Hamiltonian take the canonical form of a system of coupled scalar fields with Euclidean kinetic terms and a time-dependent mass matrix, 
\begin{widetext}
\begin{subequations}
    \begin{align}
        S_{(2)} &= \frac{1}{2} \int \mathrm{d}\tau \, \mathrm{d}^{3}\mathbf{x} \, \bigg[ \delta_{ij} V^{\prime i} V^{\prime j} - \delta_{ij} \nabla V^{i} \cdot \nabla V^{j} - \bigg( a^{2} \mathcal{M}^{2}_{ij} - \frac{a^{\prime\prime}}{a} \delta_{ij} \bigg) V^{i} V^{j} \bigg] \, , \\
        H_{(2)} &= \frac{1}{2} \int \mathrm{d}^{3}\mathbf{x} \, \bigg[ \delta_{ij} \Pi^{i} \Pi^{j} + \delta_{ij} \nabla V^{i} \cdot \nabla V^{j} + \bigg( a^{2} \mathcal{M}^{2}_{ij}-\frac{a^{\prime\prime}}{a} \delta_{ij} \bigg) V^{i} V^{j} \bigg] \, .
    \end{align}
\end{subequations}
\end{widetext}
Working in Fourier space, it is convenient to define $v^{i}(\mathbf{k},\tau) \equiv a(\tau) \Lambda^{i}_{A}(\tau) \Phi^{A}(\mathbf{k},\tau)$ so that the quadratic action and Hamiltonian become
\begin{subequations}
    \begin{align}
        S_{(2)} &= \frac{1}{2} \int \mathrm{d}\tau \, \mathrm{d}^{3} \mathbf{k} \, \big( \delta_{ij} v^{\prime i} \bar{v}^{\prime j} - \Omega^{2}_{ij} v^{i} \bar{v}^{j} \big) \, , \label{locally-flat-action-modes} \\
        H_{(2)} &= \frac{1}{2} \int \mathrm{d}^{3} \mathbf{k} \, \big( \delta_{ij} \pi^{i} \bar{\pi}^{j} + \Omega^{2}_{ij} v^{i} \bar{v}^{j}  \big) \, , \label{locally-flat-Hamilton-modes}
    \end{align}
\end{subequations}
where $\pi_{i}(\mathbf{k},\tau)$ denotes the Fourier transform of $\Pi_{i}(\mathbf{x},\tau)$. The effective frequency-squared matrix is
\begin{equation}\label{eq:eff_freq}
    \Omega^{2}_{ij}(k,\tau) \equiv \left( k^{2} - \frac{a^{\prime\prime}}{a} \right) \! \delta_{ij} + a^{2} \mathcal{M}^{2}_{ij} \, ,
\end{equation}
which is real and symmetric for all $k$ and $\tau$. The Euler-Lagrange equations then take the manifestly canonical form of coupled harmonic oscillators,
\begin{equation}\label{eq:coor_trans_eq_mov}
    v^{\prime\prime i} + \delta^{ij} \Omega^{2}_{jk} v^{k} = 0 \, .
\end{equation}

The simplification of \eqref{eq:eq_init} into canonical form highlights two complementary roles of the field redefinition. The parallel-transport condition \eqref{eq:transport_vielbein} eliminates the connection terms obtained by contracting the Christoffel symbols with the background field velocity, while the rescaling by the scale factor cancels the damping terms in conformal time. As a result, equation \eqref{eq:coor_trans_eq_mov} describes the dynamics of first-order perturbations as those of a system of coupled harmonic oscillators with time-dependent effective masses and couplings, shaped by the geometry and dynamics of the inflationary background. Importantly, this construction does not imply that all Christoffel symbols vanish in the orthonormal frame, nor that the effects of field-space curvature are removed. Rather, it eliminates the connection components that enter the covariant evolution along the background trajectory. The geometric information remains encoded in the transformed effective mass matrix, in particular through the contribution of the field-space Riemann tensor.

\subsection{Symplectic structure of mode functions}\label{subsec:sym_struc_mode_func}

With the aid of the coordinate transformations described in \subsecref{subsec:eq_motion}, 
the perturbation equations of motion can be recast as a system of $N_{\rm f}$ coupled second-order differential equations \eqref{eq:coor_trans_eq_mov}, which may be written in matrix form as
\begin{equation}\label{eq:eq-mov}
    \mathbf{v}^{\prime\prime} + \mathbf{\Omega}^{2} \mathbf{v} = \mathbf{0} \, ,
\end{equation}
here $\mathbf{v} = \begin{pmatrix} v^{1} & \cdots & v^{N_{\rm f}} \end{pmatrix}^{\intercal}$ 
is an $N_{\rm f}$-dimensional complex-valued vector, and $\mathbf{\Omega}^{2}$ is the time-dependent 
frequency-squared matrix, an $N_{\rm f} \times N_{\rm f}$ real-valued symmetric matrix whose 
entries are given by \eqref{eq:eff_freq}. 
Expressing the equations of motion in this form allows us to make explicit the symplectic structure of the dynamical system, which, as we demonstrate in this subsection, plays a central role in ensuring the preservation of the canonical commutation relations. To this end, we reduce equation \eqref{eq:eq-mov} to a system of $2N_{\rm f}$ first-order differential equations by introducing the 
phase-space vector $\mathbf{X} = \begin{pmatrix} \mathbf{v} & \mathbf{v}^{\prime} \end{pmatrix}^{\intercal}$. In terms of $\mathbf{X}$, the system takes the canonical 
Hamiltonian form
\begin{equation}\label{symp-eqn}
    \mathbf{X}^{\prime} = \mathbf{A} \mathbf{X} \, , \quad \text{with} \quad \mathbf{A} (k,\tau) \equiv \begin{pmatrix} \mathbf{0} & \mathbf{I} \\ -\mathbf{\Omega}^{2} & \mathbf{0} \end{pmatrix} \, ,
\end{equation}
where, throughout this work, $\mathbf{I}$ denotes the identity matrix and $\mathbf{0}$ the zero matrix or vector, with dimensions fixed by the context.

Since \eqref{symp-eqn} is a linear homogeneous system of dimension $2N_{\rm f}$, its general solution is a linear combination of $2N_{\rm f}$ linearly independent solutions. A convenient choice of basis consists of $N_{\rm f}$ complex solutions $\mathbf{X}_{\hat{\alpha}} = \begin{pmatrix} \mathbf{u}_{\hat{\alpha}} & \mathbf{u}_{\hat{\alpha}}^{\prime} \end{pmatrix}^{\intercal}$, with $\hat{\alpha} \in \{ 1, 2, \dots, N_{\rm f} \}$, together with their $N_{\rm f}$ complex conjugates. The corresponding integration constants may depend on the wavevector $\mathbf{k}$. Nevertheless, the basis solutions themselves, $\mathbf{X}_{\hat{\alpha}}(k,\tau)$ and $\bar{\mathbf{X}}_{\hat{\alpha}}(k,\tau)$, depend only on the magnitude $k = |\mathbf{k}|$, since the system \eqref{symp-eqn} is isotropic in $\mathbf{k}$.

The fundamental matrix of the system is obtained by assembling these $2N_{\rm f}$ independent solutions into a single  $2N_{\rm f} \times 2N_{\rm f}$ matrix. Explicitly,
\begin{equation}\label{eq:fund_mtx_ord}
    \begin{aligned}
        \mathbf{\Psi} &= \begin{pmatrix} \bar{\mathbf{X}}_{1} & \cdots & \bar{\mathbf{X}}_{N_{\rm f}} & \mathbf{X}_{1} & \cdots & \mathbf{X}_{N_{\rm f}} \end{pmatrix}
        \\
        &= \begin{pmatrix} \bar{\mathbf{u}}_{1} & \cdots & \bar{\mathbf{u}}_{N_{\rm f}} & \mathbf{u}_{1} & \cdots & \mathbf{u}_{N_{\rm f}} \\ \bar{\mathbf{u}}_{1}^{\prime} & \cdots & \bar{\mathbf{u}}_{N_{\rm f}}^{\prime} & \mathbf{u}_{1}^{\prime} & \cdots & \mathbf{u}_{N_{\rm f}}^{\prime} \end{pmatrix}
        \\
        &= \begin{pmatrix} \bar{\mathbf{U}} & \mathbf{U} \\ \bar{\mathbf{U}}^{\prime} & \mathbf{U}^{\prime} \end{pmatrix} \, ,
    \end{aligned}
\end{equation}
where $\mathbf{U} = \begin{pmatrix} \mathbf{u}_{1} & \cdots & \mathbf{u}_{N_{\rm f}} \end{pmatrix}$ is the $N_{\rm f} \times N_{\rm f}$ matrix whose columns form a basis of independent mode functions $\mathbf{u}_{\hat{\alpha}}$.

Each mode function satisfies the coupled oscillator equation $\mathbf{u}_{\hat{\alpha}}^{\prime\prime} + \mathbf{\Omega}^{2} \mathbf{u}_{\hat{\alpha}} = \mathbf{0}$ and therefore the matrix $\mathbf{U}$ obeys the compact relation $\mathbf{U}^{\prime\prime} + \mathbf{\Omega}^{2} \mathbf{U} = \mathbf{0}$. Hence, the dynamics of the full system are entirely encoded in the evolution of the basis matrix $\mathbf{U}$, whose columns span the space of solutions to \eqref{eq:eq-mov}.

By construction, the fundamental matrix $\mathbf{\Psi}$ is non-singular at all times $\tau$ for which the chosen set of solutions remains linearly independent. Moreover, it satisfies the matrix differential equation $\mathbf{\Psi}^{\prime} = \mathbf{A} \mathbf{\Psi}$, which follows directly from \eqref{symp-eqn} and the definition of its columns. This formulation provides the natural starting point for the analysis of the symplectic properties of the evolution of the operators.

The Wronskian of the system is defined as the determinant of the fundamental matrix,
\begin{equation}\label{eq:wronskian}
    W = \det \mathbf{\Psi} \, .
\end{equation}
Since $\mathbf{\Psi}^{\prime} = \mathbf{A} \mathbf{\Psi}$, Liouville’s formula gives
\begin{equation}\label{eq:wronskian-deriv}
    \frac{\mathrm{d}W}{\mathrm{d}\tau} = \mathrm{tr} (\mathbf{A}) W = 0 \, .
\end{equation}
where in the last equality we used the block structure of $\mathbf{A}$ in \eqref{symp-eqn}, which implies $\mathrm{tr}(\mathbf{A}) = 0$. Therefore, the Wronskian is conserved in time. In particular, if $\mathbf{\Psi}$ is non-singular at some initial time, it remains non-singular throughout the evolution.

The conservation of the Wronskian reflects a deeper geometric structure, namely the symplectic nature of the phase-space dynamics. To make this structure explicit, let $\alpha \in \{ 1, 2, \dots, 2N_{\rm f} \}$ label the columns of the fundamental matrix $\mathbf{\Psi}$, ordered as
\begin{equation}
    \mathbf{\Psi}_{\alpha} = \begin{cases} \bar{\mathbf{X}}_{\hat{\alpha}} \, , & \alpha = \hat{\alpha} \, , \\ \mathbf{X}_{\hat{\alpha}} \, , &  \alpha = \hat{\alpha} + N_{\rm f} \, . \end{cases}
\end{equation}
For any pair of solutions $\mathbf{\Psi}_{\alpha}$ and $\mathbf{\Psi}_{\beta}$, we define their associated symplectic bilinear form as
\begin{equation}\label{eq:symp-bilin-form}
    \omega(\mathbf{\Psi}_{\alpha} , \mathbf{\Psi}_{\beta}) \equiv \mathbf{\Psi}_{\alpha}^{\intercal} \mathbf{J} \mathbf{\Psi}_{\beta} = (\mathbf{\Psi}^{\intercal} \mathbf{J} \mathbf{\Psi})_{\alpha\beta} \, .
\end{equation}
where $\mathbf{J}$ denotes the canonical constant antisymmetric matrix.
\begin{equation}
    \mathbf{J} = \begin{pmatrix} \mathbf{0} & \mathbf{I} \\ -\mathbf{I} & \mathbf{0} \end{pmatrix} \, .
\end{equation}
Throughout this manuscript, the dimensions of $\mathbf{J}$, as well as those of the identity and zero matrices appearing in its block representation, are fixed by the context in which it acts. In the present case, $\mathbf{J}$ acts on the $2N_{\rm f}$-dimensional phase space, so that $\mathbf{I}$ and $\mathbf{0}$ are $N_{\rm f} \times N_{\rm f}$ matrices and $\mathbf{J}$ is a $2N_{\rm f} \times 2N_{\rm f}$ matrix. The bilinear form $\omega$ is antisymmetric and non-degenerate, and therefore defines a symplectic structure on the space of solutions.

The preservation of $\omega$ under time evolution follows from the fact that
$\mathbf{A}$ is a Hamiltonian matrix, i.e., a generator of the symplectic group. Concretely, $\mathbf{A}$ satisfies $\mathbf{A}^{\intercal} \mathbf{J} + \mathbf{J} \mathbf{A} = \mathbf{0}$. 
As a consequence of this,
\begin{equation}\label{eq:symp-bbilin-deriv}
    \frac{\mathrm{d}}{\mathrm{d}\tau} \omega(\mathbf{\Psi}_{\alpha} , \mathbf{\Psi}_{\beta}) = 0 \, . 
\end{equation}
Hence, $\mathbf{\Psi}^{\intercal} \mathbf{J} \mathbf{\Psi}$ is a constant of motion. Because this quantity is conserved, its value can be fixed by an appropriate choice of initial conditions. A particularly convenient normalization is
\begin{equation}\label{eq:symp_form_const_mtx_form}
    \mathbf{\Psi}^{\intercal} \mathbf{J} \mathbf{\Psi} = i \mathbf{J}\,,
\end{equation}
imposed on an initial time slice. By virtue of \eqref{eq:symp-bbilin-deriv}, this relation then holds at all times. 

When each of the individual columns is developed explicitly, this choice of initial conditions fixes the orthonormality conditions
\begin{equation}\label{eq:symp_form_const}
    \omega (\bar{\mathbf{X}}_{\hat{\alpha}} , \mathbf{X}_{\hat{\beta}}) = \bar{\mathbf{u}}_{\hat{\alpha}} \cdot \mathbf{u}_{\hat{\beta}}^{\prime} - \bar{\mathbf{u}}_{\hat{\alpha}}^{\prime} \cdot \mathbf{u}_{\hat{\beta}} = i \delta_{\hat{\alpha} \hat{\beta}} \, ,
\end{equation}
together with the vanishing of $\omega (\mathbf{X}_{\hat{\alpha}} , \mathbf{X}_{\hat{\beta}})$ and $\omega (\bar{\mathbf{X}}_{\hat{\alpha}} , \bar{\mathbf{X}}_{\hat{\beta}})$. These relations generalize the standard Wronskian normalization of a single harmonic oscillator to the coupled multifield system.

Also, the constraint in \eqref{eq:symp_form_const_mtx_form} implies that the dual relation $\mathbf{\Psi} \mathbf{J} \mathbf{\Psi}^{\intercal} = i \mathbf{J}$ also holds, which, in terms of the mode function submatrices $\mathbf{U}$, is equivalent to
\begin{equation}\label{eq:canon-conds}
    \begin{pmatrix} \bar{\mathbf{U}} \mathbf{U}^{\intercal} - \mathbf{U} \bar{\mathbf{U}}^{\intercal} & \bar{\mathbf{U}} \mathbf{U}^{\prime\intercal} - \mathbf{U} \bar{\mathbf{U}}^{\prime\intercal} \\ \bar{\mathbf{U}}^{\prime} \mathbf{U}^{\intercal} - \mathbf{U}^{\prime} \bar{\mathbf{U}}^{\intercal} & \bar{\mathbf{U}}^{\prime} \mathbf{U}^{\prime\intercal} - \mathbf{U}^{\prime} \bar{\mathbf{U}}^{\prime\intercal} \end{pmatrix} = \begin{pmatrix} \mathbf{0} & i \mathbf{I} \\ -i \mathbf{I} & \mathbf{0} \end{pmatrix} \, .
\end{equation}
These relations constitute the canonical normalization conditions for the mode functions. They ensure that the evolution, defined by $\mathbf{U}$, preserves the canonical commutation relations and is therefore consistent with the underlying symplectic structure of the classical phase space.

This enables us to rephrase the conservation of the Wronskian in terms of these symplectic bilinear forms. Explicitly, the Wronskian, as defined in \eqref{eq:wronskian}, can be written as
\begin{equation}\label{eq:Wronskian_constraint}
    \begin{aligned}
        W = \frac{\mathrm{Pf}(\mathbf{\Psi}^{\intercal} \mathbf{J} \mathbf{\Psi})}{\mathrm{Pf}(\mathbf{J})} = i^{N_{\rm f}} \, ,
    \end{aligned}
\end{equation}
where $\mathrm{Pf}(\mathbf{M})$ denotes the Pfaffian of a skew-symmetric matrix $\mathbf{M}$. It follows that the constancy of $W$ in \eqref{eq:wronskian-deriv} is  equivalent to the conservation of the symplectic form in \eqref{eq:symp-bbilin-deriv}.

With this formalism in place, the general solution of the differential equation in \eqref{eq:eq-mov} can be expressed as a linear combination of the mode functions and their complex conjugates,
\begin{equation}\label{eq:gen-sol}
    \mathbf{v}(\mathbf{k},\tau) = \sum_{\hat{\alpha}=1}^{N_{\rm f}} \Big[ c_{\hat{\alpha}}(\mathbf{k}) \bar{\mathbf{u}}_{\hat{\alpha}}(k,\tau) + \bar{c}_{\hat{\alpha}}(-\mathbf{k}) \mathbf{u}_{\hat{\alpha}}(k,\tau) \Big] \, ,
\end{equation}
where the wavevector dependence of the integration constants and the mode functions is determined in such a way that the inverse Fourier transform of $\mathbf{v}(\mathbf{k},\tau)$ renders an array of evolving real-valued scalar fields in position space.  

\subsection{Canonical quantization}\label{sec:2C}

Since the symplectic structure preserves the canonical commutation relations under mode evolution, it provides a natural framework for the canonical quantization of the field multiplet $V^{i}(\mathbf{x},\tau)$. To investigate the spectral content of the perturbation fields, we therefore transform the field operators to Fourier space,
\begin{equation}
    \hat{v}^{i}(\mathbf{k}, \tau) = \sum_{\hat{\alpha} = 1}^{N_{\rm f}} \Big[ \bar{U}^{i}_{\hat{\alpha}} (k, \tau) \hat{a}_{\hat{\alpha}}(\mathbf{k}) + U^{i}_{\hat{\alpha}} (k, \tau) \hat{a}^{\dagger}_{\hat{\alpha}} (- \mathbf{k}) \Big] \, ,
\end{equation}
which is obtained from \eqref{eq:gen-sol} by promoting the constants of integration to 
annihilation-creation operators, where the index $\hat{\alpha}$ labels the independent modes of 
vibration of the field multiplet\footnote{It is important to notice that, unlike for repeated 
field-space or spacetime indices, no Einstein summation is implied over the vibration mode 
indices $\hat{\alpha}$.}. The operators $\hat{a}_{\hat{\alpha}}$ and 
$\hat{a}^{\dagger}_{\hat{\alpha}}$ satisfy the canonical creation-annihilation algebra:
\begin{equation}
    \big[ \hat{a}_{\hat{\alpha}}(\mathbf{k}) , \hat{a}^{\dagger}_{\hat{\beta}}(\mathbf{k}^{\prime}) \big] = \delta_{\hat{\alpha}\hat{\beta}} \delta^{3} (\mathbf{k} - \mathbf{k}^{\prime}) \, .
\end{equation}
Similarly, we can express the canonical conjugate momenta $\hat{\Pi}_{i}(\mathbf{x},\tau)$ in terms of its Fourier components $\hat{\pi}_{i}(\mathbf{k},\tau)$. In Fourier mode representation, the field and canonical momenta satisfy the following 
equal-time commutation relations
\begin{equation}
    \begin{aligned}
        \big[ \hat{v}^{i}(\mathbf{k}, \tau) , \hat{\pi}^{j} (\mathbf{k}^{\prime}, \tau) \big] &= i \delta^{ij} \delta^{3} (\mathbf{k} + \mathbf{k}^{\prime})\,,
        \label{eq:can_comm_rel}
    \end{aligned}
\end{equation}
with all other commutators vanishing. 
These relations are preserved at all times by an appropriate choice of initial conditions for the mode functions, encoded in the matrix $\mathbf{U}$ with components  $U^{i}_{\hat{\alpha}}$, which is chosen to satisfy the canonical normalization conditions presented in \eqref{eq:canon-conds}.

With the Fourier expansion and the canonical commutators in place, the quantized Hamiltonian can be expressed in terms of the creation and annihilation operators. Starting from the classical Hamiltonian density \eqref{locally-flat-Hamilton-modes}, the corresponding quantized Hamiltonian is
\begin{equation}
    \hat{H} = \frac{1}{2} \int \mathrm{d}^{3} \mathbf{k} \, \big( \delta_{ij} \hat{\pi}^{i} \hat{\pi}^{\dagger j} + \Omega^{2}_{ij} \hat{v}^{i} \hat{v}^{\dagger j} \big) \, .
\end{equation}
The vacuum $|0\rangle$ is defined as the state annihilated by all $\hat{a}_{\hat{\alpha}}(\mathbf{k})$, that is, $\hat{a}_{\hat{\alpha}}(\mathbf{k}) |0\rangle = 0$, ensuring that $\langle 0 | \hat{H} | 0 \rangle$ corresponds to the minimum energy configuration consistent with the initial conditions imposed on the mode functions. In this vacuum, the Hamiltonian takes the explicit form
\begin{equation}
    \langle \hat{H} \rangle = \frac{1}{2} \delta^{3}(\mathbf{0}) \int \mathrm{d}^{3} \mathbf{k} \, \mathrm{tr} \big( \bar{\mathbf{U}}^{\prime} \mathbf{U}^{\prime\intercal} + \mathbf{\Omega}^{2} \bar{\mathbf{U}} \mathbf{U}^{\intercal} \big) \, ,
\end{equation}
where the divergent factor $\delta^{3}(\mathbf{0})$ is a manifestation of the infinite total volume of space. To define a well-posed spectral energy density, one can first place the system in a finite box of volume $V$ with periodic boundary conditions. In this case, the continuum delta function in Fourier space is replaced by the box volume, $(2\pi)^{3} \delta^{3}(\mathbf{0}) \to V$, and the total energy is written as a discrete sum over Fourier modes. In the $V \to \infty$ limit, the sum becomes an integral, and the integrand of $\int \mathrm{d}^{3}\mathbf{k}/(2\pi)^{3}$ is identified as the spectral energy density,
\begin{equation}\label{spec-energ-dens}
    \varepsilon(k,\tau) = \frac{1}{2}  \mathrm{tr} \big( \bar{\mathbf{U}}^{\prime} \mathbf{U}^{\prime \intercal} +  \mathbf{\Omega}^{2} \bar{\mathbf{U}} \mathbf{U}^{\intercal} \big) \, ,
\end{equation}
representing the contribution of each Fourier mode $k$ to the total energy density of the system. This expression highlights that the vacuum energy is determined by the sum over all modes of the field multiplet, with contributions from both the kinetic and potential parts encoded in the mode functions. It provides a direct link between the choice of initial conditions for $\mathbf{U}(k,\tau)$ and the physical energy density of the quantum field in its ground state. Moreover, due to the approximate time-translational invariance for length scales deep inside the horizon, this   accurately describes the physical state of the system.

\subsection{Amplitude-phase decomposition}\label{AP-decomposition}

As shown in \cite{GalvezGhersi:2018haa}, a well-established approach in the single-field case is to reformulate the mode function in terms of its amplitude and phase. In this context, let us consider a canonical mode function $u(k,\tau)$, which obeys the second-order equation $u^{\prime\prime} + \Omega^{2} u = 0$, where the effective frequency $\Omega^{2}(k,\tau)$ is time-dependent. Instead of evolving the complex function $u$ directly, one may parametrize the mode equation in polar form $u = r e^{i \theta}$, with $r(k,\tau)$ and $\theta(k,\tau)$ denoting the real-valued amplitude and phase, respectively. Conservation of the Wronskian associated with this second-order equation imposes a constraint on the phase evolution:
\begin{equation}\label{eq:sing_field_Wronskian}
    2 \theta^{\prime} r^{2} = 1 \, ,
\end{equation}
which fixes $\theta^{\prime}$ once the amplitude $r$ is specified. Substituting the polar form of the solution and its corresponding  Wronskian constraint \eqref{eq:sing_field_Wronskian} into the equation of motion then yields a nonlinear second-order equation governing the amplitude:
\begin{equation}\label{eq:amp-phs_decomp_single_field}
    r^{\prime\prime} + \left(\Omega^{2} - \frac{1}{4r^4}\right) r = 0 \,,
\end{equation}
which is known as the Ermakov-Pinney  equation \cite{Pinney1950TheND}. During subhorizon evolution, 
the solution of this equation is smooth and non-oscillatory, so its 
numerical solution becomes inexpensive: the integration does not need to resolve the rapid oscillations of $u$, whose frequency is large in this regime. The phase information is instead recovered from the conserved Wronskian. This substantially improves computational efficiency and numerical stability, allowing accurate evolution over many e-folds without requiring prohibitively small time steps.

Motivated by these advantages, we now extend the amplitude-phase decomposition to the multifield case. Our construction preserves the  symplectic structure of the perturbations at first-order, provides a natural generalization of the single-field framework, and retains its computational benefits in the presence of multiple coupled degrees of freedom.

To implement this extension, we decompose the fundamental matrix of mode functions, $\mathbf{U}(k,\tau)$, into amplitude and phase components. Specifically, we write
\begin{equation}\label{eq:mtx_mode_functions}
    \mathbf{U} = \begin{pmatrix} \mathbf{r}_{1} e^{i \theta_{1}} & \cdots & \mathbf{r}_{N_{\rm f}} e^{i \theta_{N_{\rm f}}} \end{pmatrix} \, ,
\end{equation}
where each column corresponds to an independent mode. Here $\mathbf{r}_{\hat{\alpha}}(k,\tau)$ is an $N_{\rm f}$-dimensional complex valued vector, rather than a real array of scalars as one might expect from a direct extension of the single-field case.  Allowing $\mathbf{r}_{\hat{\alpha}}$ to be complex provides a convenient parametrization for separating slow and fast degrees of freedom in the multifield system. The motivation for this choice and its implications will become evident in the following paragraphs. The functions $\theta_{\hat{\alpha}}(k,\tau)$ are real-valued scalar phases that introduce rapid oscillations in the subhorizon evolution of $\mathbf{u}_{\hat{\alpha}}$, with characteristic frequencies of order $k$.

Once the reparameterization of the mode solutions into amplitude and phase variables is determined, we now turn to the implications of the conserved symplectic structure for the dynamics of each mode. In particular, we will use the conservation of the canonical symplectic form to derive a constraint that relates the phase velocity $\theta^{\prime}_{\hat{\alpha}}$ to the amplitude vector $\mathbf{r}_{\hat{\alpha}}$. Let a single column of the mode function matrix be written as $\mathbf{u}_{\hat{\alpha}} = \mathbf{r}_{\hat{\alpha}} \exp (i \theta_{\hat{\alpha}})$ and define the associated phase-space vector $\mathbf{X}_{\hat{\alpha}} = \begin{pmatrix} \mathbf{u}_{\hat{\alpha}} & \mathbf{u}_{\hat{\alpha}}^{\prime} \end{pmatrix}^{\intercal}$. The canonical symplectic bilinear form evaluated on $\mathbf{X}_{\hat{\alpha}}$ and its complex conjugate $\bar{\mathbf{X}}_{\hat{\alpha}}$ is
\begin{equation}\label{symp-bil-form-AP}
    \begin{aligned}
        \omega(\bar{\mathbf{X}}_{\hat{\alpha}}, \mathbf{X}_{\hat{\alpha}}) &= 2i  \big[ \, \mathrm{Im} \, (\bar{\mathbf{r}}_{\hat{\alpha}} \cdot \mathbf{r}_{\hat{\alpha}}^{\prime}) + \theta_{\hat{\alpha}}^{\prime} r_{\hat{\alpha}}^{2} \big]
    \end{aligned}
\end{equation}
which, in accordance with the discussion  in \subsecref{subsec:sym_struc_mode_func}, is known to be conserved under arbitrary mode reparameterizations. Here $r_{\hat{\alpha}}^{2} = \mathbf{r}_{\hat{\alpha}} \cdot \bar{\mathbf{r}}_{\hat{\alpha}}$ is the squared amplitude of the mode. However, the amplitude-phase decomposition $\mathbf{u}_{\hat{\alpha}} = \mathbf{r}_{\hat{\alpha}} \exp (i \theta_{\hat{\alpha}})$ is not unique: there is a local $\mathrm{U} (1)$ gauge freedom, corresponding to the rephasing
\begin{equation}\label{AP-gauge-trans}
    \begin{aligned}
        \mathbf{r}_{\hat{\alpha}}(k,\tau) &\mapsto \tilde{\mathbf{r}}_{\hat{\alpha}}(k,\tau) = \mathbf{r}_{\hat{\alpha}}(k,\tau) \exp [i \phi_{\hat{\alpha}}(k,\tau)] \\
        \theta_{\hat{\alpha}}(k,\tau) &\mapsto \tilde{\theta}_{\hat{\alpha}}(k,\tau) = \theta_{\hat{\alpha}}(k,\tau) - \phi_{\hat{\alpha}}(k,\tau) \, .
    \end{aligned}
\end{equation}
After performing this transformation, the term $\mathrm{Im} \, (\bar{\mathbf{r}}_{\hat{\alpha}} \cdot \mathbf{r}_{\hat{\alpha}}^{\prime})$ inside the squared brackets in \eqref{symp-bil-form-AP} shifts as
\begin{equation}
    \mathrm{Im} \, (\bar{\mathbf{r}}_{\hat{\alpha}} \cdot {\mathbf{r}}_{\hat{\alpha}}^{\prime}) \mapsto \mathrm{Im} \, (\bar{\tilde{\mathbf{r}}}_{\hat{\alpha}} \cdot \tilde{\mathbf{r}}_{\hat{\alpha}}^{\prime}) = \mathrm{Im} \, (\bar{\mathbf{r}}_{\hat{\alpha}} \cdot \mathbf{r}_{\hat{\alpha}}^{\prime}) + \phi_{\hat{\alpha}}^{\prime} r_{\hat{\alpha}}^{2}\,.
\end{equation}
In consequence, it is safe to impose the following gauge choice:
\begin{equation}\label{eq:gauge-fixing}
    \phi_{\hat{\alpha}}^{\prime} =  - \frac{\mathrm{Im} \, (\bar{\mathbf{r}}_{\hat{\alpha}} \cdot \mathbf{r}_{\hat{\alpha}}^{\prime})}{r_{\hat{\alpha}}^{2}}
\end{equation}
which enforces the condition $\mathrm{Im} \, (\bar{\tilde{\mathbf{r}}}_{\hat{\alpha}} \cdot \tilde{\mathbf{r}}_{\hat{\alpha}}^{\prime}) = 0$ at all times. 
In what remains of this paper, we adopt this gauge choice and, for conciseness, drop the tilde 
notation. This gauge condition is well-defined as long as $r_{\hat{\alpha}} \neq 0$. 
Points where $r_{\hat{\alpha}} = 0$ correspond to coordinate singularities 
of the amplitude-phase parametrization rather than physical singularities, 
and can be removed by an appropriate redefinition of the solution basis, 
provided that the fundamental matrix $\bf{\Psi}$ remains non-degenerate. This is 
automatically satisfied by the conservation of a nonzero Wronskian. With these conditions 
in mind, we rewrite the conserved quantity in \eqref{symp-bil-form-AP} in the following 
simplified form
\begin{equation}\label{eq:symp_bilin_mode_func}
    \omega(\bar{\mathbf{X}}_{\hat{\alpha}} , \mathbf{X}_{\hat{\alpha}}) = 2i \theta_{\hat{\alpha}}^{\prime} r_{\hat{\alpha}}^{2} \,.
\end{equation}
Moreover, the normalization condition \eqref{eq:symp_form_const}, which arises from the canonical quantization and the underlying symplectic structure, fixes the value of this conserved quantity at all times, yielding
\begin{equation}\label{eq:AP_constraint}
    2 \theta_{\hat{\alpha}}^{\prime} r_{\hat{\alpha}}^{2} = 1 \,,
\end{equation}
where it is clear that this is analogous to the single-field constraint in 
\eqref{eq:sing_field_Wronskian}. The derivative of this 
constraint with respect to conformal time yields a  first-order 
differential equation for each mode of vibration, which 
corresponds to the conservation law for the angular momentum-like quantity in \eqref{eq:AP_constraint}:
\begin{equation}
    \theta_{\hat{\alpha}}^{\prime\prime} + \frac{(r_{\hat{\alpha}}^{2})^{\prime}}{r_{\hat{\alpha}}^{2}} \theta_{\hat{\alpha}}^{\prime} = 0\,.
\end{equation}
This relation encodes the dynamical consistency between the amplitude 
$\mathbf{r}_{\hat{\alpha}}$ and the phase velocity $\theta_{\hat{\alpha}}^{\prime}$. 
Thus, due to the conservation of the symplectic structure, the evolution of the phase is 
entirely determined by the amplitude. Furthermore, the condition \eqref{eq:gauge-fixing} 
can be understood as a gauge choice that removes the redundancy in the amplitude-phase 
decomposition. 

Once the constraint connecting the amplitude with the phase velocities is in place, we can now focus on the dynamical evolution of the amplitude vectors themselves. In analogy to the single-field case, we can substitute $\mathbf{u}_{\hat{\alpha}} = \mathbf{r}_{\hat{\alpha}} \exp (i \theta_{\hat{\alpha}})$ into the mode equation of motion $\mathbf{u}_{\hat{\alpha}}^{\prime\prime} + \mathbf{\Omega}^{2} \mathbf{u}_{\hat{\alpha}} = \mathbf{0}$ to yield the coupled system
\begin{equation}\label{EOM-AP-decomp}
    \mathbf{r}_{\hat{\alpha}}^{\prime\prime} + (\mathbf{\Omega}^{2} - \theta_{\hat{\alpha}}^{\prime 2} \mathbf{I}) \mathbf{r}_{\hat{\alpha}} + i \big[ \theta_{\hat{\alpha}}^{\prime\prime} \mathbf{r}_{\hat{\alpha}} + 2 \theta_{\hat{\alpha}}^{\prime} \mathbf{r}_{\hat{\alpha}}^{\prime} \big] = \mathbf{0} \, .
\end{equation}
The equation of motion can be equivalently expressed in terms of the real and imaginary parts of $\mathbf{r}_{\hat{\alpha}}$. To this end, we introduce the $N_{\rm f} \times 2$ real-valued matrix
\begin{equation}
    \mathbf{R}_{\hat{\alpha}} = \begin{pmatrix} \mathrm{Re} \, \mathbf{r}_{\hat{\alpha}} & \mathrm{Im} \, \mathbf{r}_{\hat{\alpha}} \end{pmatrix}
\end{equation}
so that the dynamics of $\mathbf{r}_{\hat{\alpha}}$ can be compactly recast as
%
\begin{equation}\label{eq:EOM_AP_decomp_ReIm}
    \mathbf{R}_{\hat{\alpha}}^{\prime\prime}+ \Bigg(\mathbf{\Omega}^{2} - \frac{1}{4r_{\hat{\alpha}}^4} \mathbf{I} \Bigg) \mathbf{R}_{\hat{\alpha}} + \Bigg[ \frac{1}{r_{\hat{\alpha}}^2}\mathbf{R}_{\hat{\alpha}}^{\prime} - \frac{(r_{\hat{\alpha}}^{2})^{\prime}}{2r_{\hat{\alpha}}^{4}}  \mathbf{R}_{\hat{\alpha}} \Bigg] \mathbf{J}  = \mathbf{0} \, ,
\end{equation}
%
where the $2 \times 2$ matrix $\mathbf{J}$ encodes the coupling between the real and imaginary components. Therefore, considering the relations $r_{\hat{\alpha}}^{2} = \mathrm{tr} \, \big( \mathbf{R}_{\hat{\alpha}}^{\intercal} \mathbf{R}_{\hat{\alpha}} \big)$ and $(r_{\hat{\alpha}}^{2})^{\prime} =  2 \, \mathrm{tr} \, \big( \mathbf{R}_{\hat{\alpha}}^{\intercal} \mathbf{R}_{\hat{\alpha}}^{\prime} \big)$, we have derived a set of nonlinear  
evolution equations for $\mathbf{R}_{\hat{\alpha}}$ that are 
completely decoupled from the evolving auxiliary phase variables $\theta_{\hat{\alpha}}$ and their time derivatives. 

The resulting equations of motion in \eqref{eq:EOM_AP_decomp_ReIm} are the multifield version of the Ermakov-Pinney nonlinear equation shown in \eqref{eq:amp-phs_decomp_single_field} (also in section III, eq.~(14) in \cite{GalvezGhersi:2018haa})\footnote{Setting $\mathbf{J=0}$ provides a heuristic consistency check with the amplitude equation of motion in the single-field scenario.}. In the single-field scenario, it is straightforward to show that the matrix of mode functions \eqref{eq:mtx_mode_functions} reduces to a single complex-valued function $u = r e^{i\theta}$, and the gauge-fixing condition \eqref{eq:gauge-fixing} allows $r$ to acquire an overall constant phase, $r = |r| e^{i \phi_0}$. However, since the mode equations $u^{\prime\prime} + \Omega^{2} u = 0$ form a system of second-order linear ordinary differential equations, the overall constant phase of the solution is physically irrelevant and can be absorbed by a phase redefinition $\theta (k,\tau) \mapsto \theta (k,\tau) + \phi_{0}$. Taking advantage of this freedom, one can always choose $r$ to be strictly real and positive. With this choice, the imaginary part of the mode equation \eqref{EOM-AP-decomp} vanishes identically, and the equation reduces exactly to the well-known single-field form \eqref{eq:amp-phs_decomp_single_field}. 

Up to this point, it is clear that the amplitude-phase decomposition developed here is a direct and consistent generalization of the single-field formalism. Nevertheless, when 
following the same line of reasoning used to derive \eqref{eq:amp-phs_decomp_single_field} and \eqref{eq:EOM_AP_decomp_ReIm}, this variable definition exhibits a 
fundamental difference in the multifield case. Although there is still a 
freedom to redefine phases, no gauge choice can absorb the phase freedom 
for each component of the $N_{\rm f}$-dimensional complex vectors at all 
times. This is because the effective frequency matrix $\mathbf{\Omega}^2$ 
in \eqref{EOM-AP-decomp} is, in general, non-diagonal and time-dependent. 
Due to this form of coupling, the field components of 
$\mathbf{r}_{\hat{\alpha}}$ unavoidably mix real and imaginary parts as 
the solutions evolve. Furthermore, requiring the amplitudes to be real 
overconstrains the dynamical system, as it demands an additional 
gauge-fixing condition beyond \eqref{eq:gauge-fixing}.
Thus, the amplitude vectors must be treated as genuinely complex 
objects, with their imaginary parts encoding essential information about 
the coupling of modes due to field-space curvature, potential and time 
derivatives of the background fields\footnote{Mode coupling occurs only between different field components and does not involve the vibration modes. Solutions with different values of $\hat{\alpha}$ remain decoupled.}.

\subsection{Two-point correlation functions, power spectrum and phase-space observables}\label{subsec:two_pt_pow_spec}

The assumption of primordial Gaussianity implies that the statistical properties of inflationary perturbations are completely characterized by their two-point correlation functions in Fourier space. In the multifield case, it is convenient to treat the curvature and isocurvature perturbations on equal footing by introducing the collective set of operators $\hat{\mathcal{X}} \in \{ \hat{\mathcal{R}}, \hat{\mathcal{S}}_{1}, \hat{\mathcal{S}}_{2}, \dots, \hat{\mathcal{S}}_{N_{\rm f}-1} \}$. These operators correspond to projections of the field perturbations along a time-dependent orthonormal basis in field space consisting of the adiabatic direction and the $N_{\rm f} - 1$ independent entropic directions orthogonal to it.

In this notation, the Fourier modes of the perturbation operators are related to the canonical variables $\hat{v}^{i}(\mathbf{k},\tau)$ through
\begin{equation}
    \hat{\mathcal{X}}(\mathbf{k},\tau) = \frac{H}{|\dot{\varphi}|} e_{\cal X}^{A} \hat{\Phi}_{A} = \frac{H}{a |\dot{\varphi}|} e_{\cal X}^{A} h_{AB} \Lambda^{B}_{i} \hat{v}^{i} \, .
\end{equation}
Here, the vectors $e_{\mathcal{X}}^{A}$, with $\mathcal{X}\in\{\mathcal{R},\mathcal{S}_{1},\ldots, \mathcal{S}_{N_{\rm f}-1}\}$, form the orthonormal field-space basis adapted to the background trajectory introduced in \subsecref{subsec:back}. The vector $e_{\mathcal{R}}^{A}$ is tangent to the trajectory, while the vectors $e_{\mathcal{S}_{a}}^{A}$ span the subspace orthogonal to it. Accordingly, $\hat{\mathcal{R}}$ represents the adiabatic perturbation along the background motion, while $\hat{\mathcal{S}}_{a}$ denote the independent isocurvature perturbations orthogonal to it.

The statistical properties of these perturbations are encoded in their equal-time two-point correlation functions. For any pair of operators $\hat{\mathcal{X}}$ and $\hat{\mathcal{Y}}$, the corresponding dimensionless power spectra are defined through
\begin{equation}
    \langle \hat{\mathcal{X}}(\mathbf{k},\tau) \hat{\mathcal{Y}}(\mathbf{k}^{\prime},\tau) \rangle = \frac{2\pi^{2}}{k^{3}} \mathcal{P}_{\cal XY} (k,\tau) \, \delta^{3}(\mathbf{k} + \mathbf{k}^{\prime}) \, .
\end{equation}
The diagonal components describe the individual variances of the curvature and isocurvature modes, while the off-diagonal components encode correlations between them. In practice, it is useful to consider the total curvature-isocurvature power of cross-correlations and the total isocurvature power:
\begin{equation}
    \mathcal{P}_{\cal RS} \equiv \sqrt{ \sum_{a=1}^{N_{\rm f}-1} \mathcal{P}_{\mathcal{R} \mathcal{S}_{a}}^{2} } \, , \quad \mathcal{P}_{\cal SS} \equiv \sum_{a=1}^{N_{\rm f}-1} \mathcal{P}_{\mathcal{S}_{a} \mathcal{S}_{a}} \, .
\end{equation}
These quantities are invariant under rotations of the isocurvature basis. This 
follows from the fact that the vectors $e_{\mathcal{S}_{a}}^{A}$ are not uniquely 
defined, since any orthogonal transformation within the $(N_{\rm f}-1)$-dimensional 
entropic subspace yields an equally valid basis. Under such transformations, the 
components $\mathcal{P}_{\mathcal{R}\mathcal{S}_{a}}$ transform as a vector, so that 
$\mathcal{P}_{\cal RS}$ is its invariant Euclidean norm, while the spectra 
$\mathcal{P}_{\mathcal{S}_{a}\mathcal{S}_{b}}$ transform as the components of the 
isocurvature block of the power-spectrum matrix, whose trace is 
$\mathcal{P}_{\cal SS}$ and is likewise invariant.

Substituting the relation between $\hat{\mathcal{X}}$ and the canonical perturbations into the two-point function yields the general expression
\begin{equation}
    \mathcal{P}_{\mathcal{X}\mathcal{Y}} = \frac{k^{3}}{2\pi^{2}} \left( \frac{H}{a |\dot{\varphi}|} \right)^{2} e_{\cal X}^{A} h_{AB} \Lambda^{B}_{i} \big( \bar{\mathbf{U}} \mathbf{U}^{\intercal} \big)^{ij} \Lambda^{D}_{j} h_{CD} e_{\cal Y}^{C} \label{eq:def_PRR} \,,
\end{equation}
where $\big( \bar{\mathbf{U}} \mathbf{U}^{\intercal} \big)^{ij} = \sum_{\hat{\alpha}=1}^{N_{\rm f}} \bar{U}^{i}_{\hat{\alpha}} U^{j}_{\hat{\alpha}}$.

To make the notation consistent with the output variables of the evolution equations in \eqref{eq:EOM_AP_decomp_ReIm}, it is necessary to express the matrix product $\bar{\mathbf{U}} \mathbf{U}^{\intercal}$ in terms of the amplitude vectors $\mathbf{r}_{\hat{\alpha}}$ introduced in the amplitude-phase decomposition. Using the canonical normalization conditions \eqref{eq:canon-conds}, one finds that the antisymmetric part cancels, and the result simplifies to
\begin{equation}
    \bar{\mathbf{U}} \mathbf{U}^{\intercal} = \frac{1}{2} \sum_{\hat{\alpha}=1}^{N_{\rm f}} (\bar{\mathbf{r}}_{\hat{\alpha}} \mathbf{r}_{\hat{\alpha}}^{\intercal} + \mathbf{r}_{\hat{\alpha}} \bar{\mathbf{r}}_{\hat{\alpha}}^{\intercal}) = \sum_{\hat{\alpha} = 1}^{N_{\rm f}} \mathbf{R}_{\hat{\alpha}} \mathbf{R}_{\hat{\alpha}}^{\intercal} \, ,
\end{equation}
This decomposition makes it explicit that only the symmetric, real combinations of the amplitude vectors contribute to the curvature power spectrum. Consequently, all spectra $\mathcal{P}_{\cal XY}$ (including the curvature spectrum, the individual isocurvature spectra, and the curvature-isocurvature cross spectra) are manifestly real, as required by their interpretation as variances and covariances of the primordial perturbations.

The quantum state of linear perturbations during inflation admits a natural description in phase space in terms of its Wigner function, which defines a quasi-probability distribution over the canonical variables associated with the perturbations. In the present context, these variables are the field fluctuations and their conjugate momenta, denoted collectively by $\hat{v}^{i}$ and $\hat{\pi}^{i}$. For Gaussian quantum states, which include the Bunch-Davies vacuum and its subsequent evolution under linear equations of motion, the Wigner function provides a complete description of the density operator and is entirely determined by the first and second moments of these canonical field operators. The corresponding constant-value contours of the Wigner function form ellipsoids in the full phase space. In two-dimensional projections, these contours reduce to ellipses, which we refer to as Wigner ellipses.

The geometry of the Wigner ellipsoid encodes the physical properties of the quantum state. Its orientation reflects correlations among the canonical variables, while its principal axes characterize the degree and direction of squeezing. The full phase-space volume measures the quantum uncertainty and is constrained by the uncertainty principle. Under symplectic evolution, this volume is conserved, although the areas of individual two-dimensional projections may vary as correlations are redistributed among the canonical variables. These properties are particularly relevant in inflationary cosmology, where squeezing and the development of classical correlations play a central role in the quantum-to-classical transition of primordial perturbations.

A central object in this formulation is the covariance matrix, which provides a symplectically covariant characterization of the second moments of the canonical variables and therefore determines the geometry of the Wigner ellipsoid. It offers a compact description of quantum correlations, squeezing, and phase-space-volume preservation, while evolving deterministically under the linear perturbation equations. Since an initially Gaussian state remains Gaussian under linear evolution, its quantum state is completely specified by the covariance matrix together with its first moments. In the present case, where the first moments vanish, it is therefore sufficient to follow the evolution of the covariance matrix, which parameterizes the density operator through the Wigner function.

To formulate this description explicitly, it is convenient to work with a set of real canonical variables constructed from the complex field perturbations $\hat{v}^{i}$ and their conjugate momenta $\hat{\pi}^{i}$. This is achieved by separating each operator into its real and imaginary parts. Restricting wavevectors to a common half of Fourier space containing one representative of each nonzero pair $\{ \mathbf{k} , -\mathbf{k} \}$, we define, for each field-space index $i$,
\begin{equation}
    \begin{aligned}
        \hat{v}^{i}_{\rm R} \equiv \frac{\hat{v}^{i} + \hat{v}^{\dagger i}}{\sqrt{2}} \, , \quad \hat{\pi}^{i}_{\rm R} \equiv \frac{\hat{\pi}^{i} + \hat{\pi}^{\dagger i}}{\sqrt{2}} \, , \\ \hat{v}^{i}_{\rm I} \equiv \frac{\hat{v}^{i} - \hat{v}^{\dagger i}}{\sqrt{2}i} \, , \quad \hat{\pi}^{i}_{\rm I} \equiv \frac{\hat{\pi}^{i} - \hat{\pi}^{\dagger i}}{\sqrt{2}i} \, ,
    \end{aligned}
\end{equation}
where the label $\mathrm{S} \in \{ \rm R, I \}$ is used to distinguish the real and imaginary parts of the field operators. These operators are Hermitian and together form a complete set of real phase space coordinates. They obey canonical equal-time commutation relations and provide a convenient basis for constructing the covariance matrix. In the following, we use these variables to define the two-point correlation matrices and the associated covariance matrix.

The covariance matrix $\mathbf{\Sigma}(k,\tau)$ is then defined as the real $2N_{\rm f} \times 2N_{\rm f}$ matrix composed of the equal-time expectation values of anticommutators of these canonical variables. It can be written in block form as
\begin{equation}
    \mathbf{\Sigma} = \begin{pmatrix} \mathbf{\Sigma}_{vv} & \mathbf{\Sigma}_{v \pi} \\ \mathbf{\Sigma}_{\pi v} & \mathbf{\Sigma}_{\pi \pi} \end{pmatrix} \, ,
\end{equation}
where each block $\mathbf{\Sigma}_{xy}(k,\tau)$ is an $N_{\rm f} \times N_{\rm f}$ matrix. The individual blocks are defined through the two-point anticommutators
\begin{equation}
    \Big\langle \Big\{ \hat{x}_{\rm S}^{i} (\mathbf{k},\tau) , \hat{y}_{\rm S^{\prime}}^{j} (\mathbf{k}^{\prime}, \tau) \Big\} \Big\rangle = \Sigma_{xy}^{ij} (k,\tau) \, \delta_{\rm SS^{\prime}} \delta^{3}(\mathbf{k} - \mathbf{k}^{\prime}) \, ,
\end{equation}
where $\hat{x}_{\rm S}^{i}, \hat{y}_{\rm S}^{i} \in \{ \hat{v}_{\rm S}^{i} , \hat{\pi}_{\rm S}^{i} \}$ denote generic phase-space operators. The Kronecker delta $\delta_{\rm SS^{\prime}}$ reflects the statistical independence of the real and imaginary sectors.

By expressing the field and momentum operators in terms of the mode functions, the equal-time anticommutators that define the covariance matrix can be explicitly evaluated. In this way, the covariance matrix admits a compact representation in terms of the fundamental solution matrix $\mathbf{\Psi}$, introduced in \eqref{eq:fund_mtx_ord}, namely
\begin{equation}\label{eq:cov-mtx}
    \begin{aligned}
        \mathbf{\Sigma} &= \begin{pmatrix} \bar{\mathbf{U}} \mathbf{U}^{\intercal} + \mathbf{U} \bar{\mathbf{U}}^{\intercal} & \bar{\mathbf{U}} \mathbf{U}^{\prime\intercal} + \mathbf{U} \bar{\mathbf{U}}^{\prime\intercal} \\ \bar{\mathbf{U}}^{\prime} \mathbf{U}^{\intercal} + \mathbf{U}^{\prime} \bar{\mathbf{U}}^{\intercal} & \bar{\mathbf{U}}^{\prime} \mathbf{U}^{\prime\intercal} + \mathbf{U}^{\prime} \bar{\mathbf{U}}^{\prime\intercal}  \end{pmatrix} \\
        &= \mathbf{\Psi} \begin{pmatrix} \mathbf{0} & \mathbf{I} \\ \mathbf{I} & \mathbf{0} \end{pmatrix} \mathbf{\Psi}^{\intercal} \, .    
    \end{aligned}
\end{equation}
This expression makes explicit how the statistical properties of the Gaussian state are entirely determined by the classical evolution of the mode functions encoded in $\mathbf{\Psi}$.

A direct and significant consequence of \eqref{eq:cov-mtx} is that the determinant of the covariance matrix is entirely fixed by the conserved Wronskian $W$ associated with the symplectic evolution of the system. In particular, employing the Wronskian constraint relation in \eqref{eq:Wronskian_constraint}, one obtains
\begin{equation}
    \det\mathbf{\Sigma} = (-1)^{N_{\rm f}} W^{2} = 1\,,
    \label{eq:cons_area}
\end{equation}
which is independent of conformal time. Its conservation follows from the symplectic structure of the equations of motion: Hamiltonian evolution preserves the symplectic form and, consequently, the phase-space volume. Thus, the constancy of $\det\mathbf{\Sigma}$ provides a statistical manifestation of Liouville’s theorem.

Finally, by expressing each mode function in amplitude-phase form and using the real $N_{\rm f} \times 2$ matrices $\mathbf{R}_{\hat{\alpha}} = \begin{pmatrix} \mathrm{Re} \, \mathbf{r}_{\hat{\alpha}} & \mathrm{Im} \, \mathbf{r}_{\hat{\alpha}} \end{pmatrix}$, the covariance matrix acquires a fully explicit representation in terms of the mode amplitudes. In block form, one obtains
\begin{widetext}
    \begin{equation}
        \begin{aligned}
        \mathbf{\Sigma}_{vv} &= 2 \sum_{\hat{\alpha}=1}^{N_{\rm f}} \mathbf{R}_{\hat{\alpha}} \mathbf{R}_{\hat{\alpha}}^{\intercal} \, , \quad
        &
        \mathbf{\Sigma}_{v \pi} &= 2 \sum_{\hat{\alpha}=1}^{N_{\rm f}} \bigg( \mathbf{R}_{\hat{\alpha}} \mathbf{R}_{\hat{\alpha}}^{\prime\intercal} - \frac{1}{2r_{\hat{\alpha}}^2} \mathbf{R}_{\hat{\alpha}} \mathbf{J} \mathbf{R}_{\hat{\alpha}}^{\intercal} \bigg) \, ,
        \\
        \mathbf{\Sigma}_{\pi v} &= 2 \sum_{\hat{\alpha}=1}^{N_{\rm f}} \bigg( \mathbf{R}_{\hat{\alpha}}^{\prime} \mathbf{R}_{\hat{\alpha}}^{\intercal} + \frac{1}{2r_{\hat{\alpha}}^2} \mathbf{R}_{\hat{\alpha}} \mathbf{J} \mathbf{R}_{\hat{\alpha}}^{\intercal} \bigg) \, , \quad
        &
        \mathbf{\Sigma}_{\pi\pi} &= 2 \sum_{\hat{\alpha}=1}^{N_{\rm f}} \bigg[ \mathbf{R}_{\hat{\alpha}}^{\prime} \mathbf{R}_{\hat{\alpha}}^{\prime\intercal} - \frac{1}{2r_{\hat{\alpha}}^2} \big( \mathbf{R}_{\hat{\alpha}}^{\prime} \mathbf{J} \mathbf{R}_{\hat{\alpha}}^{\intercal} - \mathbf{R}_{\hat{\alpha}} \mathbf{J} \mathbf{R}_{\hat{\alpha}}^{\prime\intercal} \big) + \frac{1}{4r_{\hat{\alpha}}^4} \mathbf{R}_{\hat{\alpha}} \mathbf{R}_{\hat{\alpha}}^{\intercal} \bigg] \, .
    \end{aligned}
    \label{eq:cov_r}
\end{equation}
\end{widetext}
These expressions display explicitly how each block of the covariance matrix is constructed from quadratic combinations of the real amplitude matrices $\mathbf{R}_{\hat{\alpha}}$ and their time derivatives. This formulation is especially useful for practical applications. Analytically, it allows one to identify the separate contributions of amplitude growth and phase rotation to each covariance block. Numerically, it provides a stable and transparent parametrization in terms of real quantities.

\subsection{Initial conditions and mode injection scheme}
\label{subsec:init_cond_schm}

Having derived \eqref{eq:EOM_AP_decomp_ReIm} as the equations of motion, it is 
necessary to complete the dynamical system by specifying a convenient set of 
initial conditions. We first fix the initial conditions for the vielbein 
$\Lambda_{i}^{A}$, which defines a local orthonormal frame in field space via the metric relation \eqref{eq:old-to-new-metric}. At $\tau = \tau_{0}$, we choose the first basis vector to align with the tangent to the background trajectory, while the remaining $(N_{\rm f}-1)$ vectors span the subspace orthogonal to it. These orthogonal vectors are constructed through the Gram-Schmidt procedure with respect to the field-space metric $h_{AB}$ evaluated at $\tau_{0}$, thereby ensuring the orthonormality condition \eqref{eq:old-to-new-metric}. The subsequent evolution of the vielbein is determined by the parallel transport equation \eqref{eq:transport_vielbein}.

Any alternative initial specification of the vielbein related to this choice by a local orthogonal transformation in field space also satisfies \eqref{eq:old-to-new-metric} and defines an equally valid orthonormal frame. Thus, the construction is not unique. While the initial frame has a direct dynamical interpretation (with the first vector tangent to the trajectory and the remaining vectors orthogonal to it), this property holds strictly at $\tau = \tau_{0}$. At later times, parallel transport preserves orthonormality but does not, in general, maintain alignment with the instantaneous trajectory or the perpendicular subspace.

Once we fixed the initial conditions for the orthonormal frame in field space, we now turn to the initial conditions for the mode functions themselves within the amplitude-phase decomposition. In this representation, the spectral energy density \eqref{spec-energ-dens} can be recast as
\begin{equation}
    \varepsilon(k,\tau) = \frac{1}{2} \sum_{\hat{\alpha}=1}^{N_{\rm f}} \big( \mathbf{r}_{\hat{\alpha}}^{\prime} \cdot \bar{\mathbf{r}}_{\hat{\alpha}}^{\prime} + \theta_{\hat{\alpha}}^{\prime 2} r_{\hat{\alpha}}^{2} + \mathbf{r}_{\hat{\alpha}}^{\intercal} \mathbf{\Omega}^{2} \bar{\mathbf{r}}_{\hat{\alpha}} \big) \, .
\end{equation}
To identify the vacuum or minimal-energy configuration, we choose initial conditions that minimize the energy density at the initial time $\tau_{0}$. Since the frequency-squared matrix $\mathbf{\Omega}^{2}(k,\tau)$ is real and symmetric, it admits a complete set of orthonormal eigenvectors. Therefore, there exists an orthogonal transformation $\mathbf{P}(k,\tau) \in O(N_{\rm f})$ which diagonalizes $\mathbf{\Omega}^{2}$:
\begin{equation}
    \mathbf{P}^{\intercal} \mathbf{\Omega}^{2} \mathbf{P} = \mathrm{diag} \big( \omega_{1}^{2} , \dots , \omega_{N_{\rm f}}^{2} \big) \, .
\end{equation}
By rotating to the eigenbasis defined by $\mathbf{P}$, one can decouple the quadratic form in the energy density at the initial time and thereby identify the configuration of $\mathbf{r}_{\hat{\alpha}}$ and $\theta_{\hat{\alpha}}^{\prime}$ that yields the minimal-energy state. This construction provides the generalized vacuum initial conditions when the frequency matrix is not diagonal.

Since the diagonalization basis of $\mathbf{\Omega}^{2}$ varies with time, its time derivative introduces a connection term $\boldsymbol{\chi} = \mathbf{P}^{\intercal} \mathbf{P}^{\prime}$ that enters the amplitude dynamics. Explicitly, the mode functions in the original basis and in the instantaneous eigenbasis of $\mathbf{\Omega}^{2}$ are related by
\begin{equation}\label{eq:change-of-basis}
    \mathbf{r}_{\hat{\alpha}} = \mathbf{P} \tilde{\mathbf{r}}_{\hat{\alpha}} \, , \qquad \mathbf{r}_{\hat{\alpha}}^{\prime} = \mathbf{P}^{\prime} \tilde{\mathbf{r}}_{\hat{\alpha}} + \mathbf{P} \tilde{\mathbf{r}}_{\hat{\alpha}}^{\prime} \, .
\end{equation}
Here, $\tilde{\mathbf{r}}_{\hat{\alpha}}$ denotes the complex amplitude in the eigenbasis, while $\mathbf{r}_{\hat{\alpha}}$ is expressed in the original basis. A natural choice for the minimal-energy initial conditions in the eigenbasis is
\begin{equation}
    \tilde{\mathbf{r}}_{\hat{\alpha}} \big|_{\tau_{0}} = \frac{\mathbf{e}_{\hat{\alpha}}}{\sqrt{2 \omega_{\hat{\alpha}}}} \bigg|_{\tau_{0}} \, , \qquad \tilde{\mathbf{r}}_{\hat{\alpha}}^{\prime} \big|_{\tau_{0}} = - \boldsymbol{\chi} \tilde{\mathbf{r}}_{\hat{\alpha}} \big|_{\tau_{0}} \, ,
\end{equation}
where $\omega_{\hat{\alpha}} |_{\tau_{0}}$ is the square root of the $\hat{\alpha}$-th eigenvalue of $\mathbf{\Omega}^{2} |_{\tau_{0}}$, and $\mathbf{e}_{\hat{\alpha}}$ is the standard canonical unit vector, having a $1$ in the $\hat{\alpha}$-th position and zeros in all other entries. In the original basis, these conditions become
\begin{equation}
    \mathbf{r}_{\hat{\alpha}} \big|_{\tau_{0}} = \frac{\mathbf{P} \mathbf{e}_{\hat{\alpha}}}{\sqrt{2\omega_{\hat{\alpha}}}} \bigg|_{\tau_{0}} \, , \quad \mathbf{r}_{\hat{\alpha}}^{\prime} \big|_{\tau_{0}} = \mathbf{0} \, .
\end{equation}
With this choice, the amplitude-phase decomposition constraint \eqref{eq:AP_constraint} fixes the initial phase velocities uniquely:
\begin{equation}
    \theta_{\hat{\alpha}}^{\prime} \big|_{\tau_{0}} = \omega_{\hat{\alpha}} \big|_{\tau_{0}} \ .
\end{equation}
This choice corresponds to initializing each mode in a harmonic oscillator ground state along the principal directions of the instantaneous frequency-squared matrix $\mathbf{\Omega}^{2}$. Consequently, the spectral energy density at $\tau_{0}$ is
\begin{equation}
    \varepsilon(k,\tau_{0}) = \frac{1}{2} \sum_{\hat{\alpha}=1}^{N_{\rm f}} \omega _{\hat{\alpha}} \big|_{\tau_{0}} \, ,
\end{equation}
with each mode contributing half of its instantaneous frequency, consistent with the zero-point energy along the instantaneous eigenvectors of $\mathbf{\Omega}^{2}$.

\injscheme

Time-translation invariance of the Minkowski vacuum can be 
exploited to construct an efficient scheme for evolving the set 
of modes required to compute field observables. The solid red 
curve in \Figref{fig:inj_scheme} shows the evolution of the 
inverse Hubble scale during an inflationary phase driven by the 
potential in \eqref{eq:potential_st}. 
Potential parameters are 
identical to those used to generate the potential surface plot in 
the right panel of \Figref{fig:def_geom_V_traj} in 
\subsecref{subsec:back}. The inset at the upper left corner of 
this figure confirms that sharp turns in the background 
trajectory evolution modify the horizon evolution. However, with respect to 
$1/H$, these modifications allow us to define a constant-wavelength surface that 
serves as the starting point for the evolution of the modes. 
As the comoving wavenumber $k$ increases, fluctuation modes effectively 
scan the time-dependent inflationary horizon, probing different 
stages of its evolution. As a result, horizon deformations 
induced by sharp turns in the background trajectory are imprinted 
as features in the primordial spectra, since modes exit the 
horizon slightly earlier or later than in the quasi-de Sitter 
case, where the horizon is nearly constant.

Each mode labeled by its comoving wavenumber $k$ is injected a fixed number of e-folds after the onset of the 
background evolution. Modes with larger values of $k$ are injected at later times. This sequential initialization 
contributes significantly to the computational efficiency of the scheme when computing the spectra as functions of $k$, as it avoids unnecessarily long integrations 
for short-wavelength modes and thereby reduces the total evolution time. In the next section, we present an additional 
source of efficiency: the reduction of the effective oscillation frequency of the system. This frequency suppression 
allows for larger integration time steps, reduces numerical stiffness, and induces a further decrease in the overall computational runtime.

\section{Results}
\label{sec:results}

Thus far, we have presented the equations of motion required to evolve perturbative inhomogeneities to the first order in perturbation theory. Equations 
\eqref{eq:back_field}, \eqref{eq:hubble}, and \eqref{eq:h_dot} determine the nonlinear evolution of the background fields and the expansion history. In addition, 
solving for field perturbations in a curved field-space geometry requires evolving the vielbein according to the parallel-transport equations in 
\eqref{eq:transport_vielbein}. With these elements in place, we show that the equations of motion in \eqref{eq:EOM_AP_decomp_ReIm}, derived from the conservation 
of the system’s symplectic structure, are almost free from fast oscillatory 
behavior and dynamical instabilities. \Figref{fig:cholesky_problem} illustrates 
the onset of dynamical instabilities in the correlator-transport formulation of 
\cite{GalvezGhersi:2016wbu} in the presence of rapid changes in the background 
trajectory. The scenarios considered in this section probe this same regime, 
where such instabilities are expected to arise in that approach. It is important 
to remark that the goal is to demonstrate that the method reliably captures the 
evolution of fluctuations in nontrivial backgrounds, rather than to provide a 
detailed physical interpretation of the resulting spectral features.

\subsection{Mode evolution and the amplitude-phase decomposition}
\label{subsec:res_decomp}

\ctchecks

Up to this point, we have discussed the limitations of the 
Cholesky decomposition scheme proposed in 
\cite{GalvezGhersi:2016wbu} and introduced a dynamical framework 
that implements an amplitude-phase decomposition. This framework 
generalizes the single-field approach developed in 
\cite{GalvezGhersi:2018haa} to the case of multiple fields in a 
nontrivial field-space geometry. In this subsection, we present 
explicit results obtained from the system of equations in 
\eqref{eq:EOM_AP_decomp_ReIm}. As a first step, it is crucial to show the 
suppression of the fastest oscillation scales in the system to validate the consistency 
of the proposed method. The left panel of \Figref{fig:freq_full_ps} illustrates the reduction 
of the eigenvalues of the effective oscillation frequency matrix in \eqref{eq:EOM_AP_decomp_ReIm}, 
relative to those of the oscillation matrix $\mathbf{\Omega}^2$ in the standard mode equations. We consider 
the same two-field nonlinear potential plotted in the left panel of \Figref{fig:def_geom_V_traj}. 
In this case, the solid blue region 
contains four 
eigenvalue evolution curves: two associated with the vibration 
modes labeled by $\hat{\alpha}$ (keeping constant $i$) and two corresponding to the 
field components labeled by $i$ (for constant $\hat{\alpha}$). As for the thin red region, it contains the two eigenvalues associated 
to $\mathbf{\Omega}^2$. This behavior is consistent with 
the suppression of the effective oscillation frequency in several 
orders of magnitude, as we can note in the region shaded in green. 
This panel illustrates a distinctive feature of our 
approach also discussed in \subsecref{subsec:eq_motion}: rather than evolving the 
two-point correlators directly, we recast the mode 
equations into a nonlinear equation for the amplitude variables, whose effective 
oscillation frequency is suppressed relative to the original mode equations, allowing 
perturbations to be resolved on timescales comparable to the background evolution, 
including across sharp-turn events. This constitutes an improvement over both our 
previous scheme \cite{GalvezGhersi:2016wbu}, which instead evolved the Cholesky 
factors of the correlation matrix, and correlator-transport approaches such as 
\cite{Mulryne:2016mzv,Pinol:2023oux}, which evolve the two-point correlation matrix 
directly and, being bilinear in the mode functions, carry twice the oscillation 
frequency of the original mode equations. Although not shown here, we tested this 
frequency reduction for more abrupt background field evolutions, finding similar results.

It is also important to verify that the method 
yields results consistent with scenarios of 
smooth background evolution. To this end, we 
show in the right panel of 
\Figref{fig:freq_full_ps} that the 
amplitude-phase decomposition approach 
reproduces the same primordial curvature power 
spectrum as the Cholesky decomposition technique 
presented in \cite{GalvezGhersi:2016wbu} for a 
range of field-coupling values. Similar agreement is observed 
for the isocurvature power spectrum and its 
cross-correlations with the adiabatic modes. 
This comparison provides indirect 
confirmation of the absence of dynamical or 
numerical instabilities during the evolution of 
all $k$ modes contributing to these spectra. 

Although the agreement observed in scenarios of smooth background 
evolution provides an important validation of the method, it does 
not address its performance in more challenging regimes. The main 
motivation for this work arises from the emergence of dynamical 
instabilities in the Cholesky decomposition scheme when the 
background fields undergo abrupt changes. Hence, the objective of the 
remainder of this subsection is to demonstrate that, 
in contrast to the Cholesky decomposition approach, the 
amplitude-phase decomposition approach presented in this 
manuscript does not introduce such instabilities during mode 
evolution, even in the presence of sharp features in the 
background field evolution arising from potential deformations 
or from nontrivial field-space geometry effects. 

The parametrization of the field-space geometry in 
\eqref{eq:fld_met_ex} and of the potential in 
\eqref{eq:potential_st}, in terms of the deformation functions 
$\Delta h_{AB}$ and $\Delta V$ provides a systematic 
framework for introducing 
controlled departures from the standard configuration. This 
setup allows us to gradually switch on the effects of sharp 
turns in the background trajectory and to isolate their impact 
on the evolution of adiabatic and isocurvature scalar 
perturbations. To recall their respective roles, the 
potential deformation function $\Delta V$ parameterizes 
the controlled departures from a quadratic 
smooth potential, introducing localized features in the scalar potential. The geometric deformation
function $\Delta h_{AB}$, on the other hand, modifies the 
field-space metric through a sequence of Gaussian bumps, thereby 
generating rapid variations in the background trajectory and 
departures from flat geometry. Although the specific examples 
developed so far, namely, the deformed potential and 
field-space geometry described in \secref{sec:method}, 
are not associated with a particular physical realization, 
they nonetheless provide a sufficiently general and controllable setting in which the effects of increasingly 
sharp background turns on the perturbation modes can be systematically assessed.

\modevol

We begin by considering the case in which sharp turns are 
generated purely by geometric deformations.
The upper panels of \Figref{fig:mode_evol} illustrate the 
effects of sharp turns in the background field trajectory, 
induced by geometric deformations that drive the field-space 
metric away from flatness, on the evolution of the adiabatic, 
cross-correlation, and isocurvature power spectra at an 
arbitrary value of $k$. The potential used here is the same 
smooth two-field quartic potential employed to obtain the 
results shown in \Figref{fig:freq_full_ps}. In that case, the potential does not introduce 
sharp features in the background field evolution. From these panels, it is 
evident that the abrupt 
features in the mode evolution correlate with the sequence of 
sharp Gaussian bumps that deform the field-space geometry. 
Moreover, the inset in the upper-left panel shows that the 
evolution of the adiabatic modes is not necessarily strongly 
affected, even in cases where deviations from flat geometry are 
large enough for the cross-correlation power to become 
comparable to the adiabatic and isocurvature components. Most 
importantly, the uninterrupted evolution displayed 
across all panels demonstrates that the amplitude-phase 
decomposition formalism consistently resolves the perturbation 
dynamics, even in the presence of progressively sharper 
geometric deformations and the associated rapid turns in the 
background trajectory. 

We next consider the complementary case in which sharp features 
arise from deformations of the scalar potential.
The three lower panels of \Figref{fig:mode_evol} show the 
evolution corresponding to the cases with 
$\Delta V \ne 0$. To isolate the effects of potential 
deformations, we set $\Delta h_{AB}=0$, so that the field-space 
geometry is flat. As in the case of geometric deformations, 
the oscillatory behavior observed in the mode amplitudes and 
cross-correlators originates from turns in the background field 
trajectory. In the present case, these turns are induced by 
localized structures in the scalar potential that are similar, 
though intentionally slightly smoother, to those depicted in the 
right panel of \Figref{fig:def_geom_V_traj}. The additional 
smoothing is introduced to prevent the generation of excessively 
large power in the curvature perturbations. Deviations from the 
standard configuration (here corresponding to the two-field 
quadratic potential) are shown in progressively darker colors 
as $\Delta V$ increases. In addition to this, none of these 
panels exhibits signs of dynamical instabilities that interrupt 
the evolution, even in the cases of large power. In contrast to the upper panels, the magnitude of 
these deviations is apparent across all power spectra. A major advantage 
of the amplitude-phase decomposition approach is that the time step required 
to resolve the perturbation dynamics remains comparable to that used for the 
background evolution, and in some cases the two can even be identical.

\psk

Having analyzed the impact of the deformations on the mode evolution, we 
now turn to their imprint on the adiabatic, cross-correlation, and 
isocurvature power spectra as functions of $k$. \Figref{fig:ps_k} shows 
various features in the power spectra for the same cases of gradual 
geometric and potential deformations displayed in \Figref{fig:mode_evol}. 
As expected, continuous departures in the potential and in the 
field-space geometry manifest as corresponding deviations from the 
reference scenarios of a smooth potential and a flat metric, respectively. 

The upper panels in \Figref{fig:ps_k} illustrate the cases with geometric 
deformations $(\Delta h_{AB}\neq 0)$. In 
this setup, the total number of e-folds generated by the background 
trajectory is approximately insensitive to the magnitude of the 
departures from flat field-space geometry. Consequently, 
the end-of-inflation screen (collecting the mode amplitudes depicted in 
\Figref{fig:inj_scheme}) can be fixed in the same location for all 
deformation amplitudes. In agreement with the results shown in 
\Figref{fig:mode_evol}, the spectrum of cross-correlations can reach 
amplitudes comparable to those of the adiabatic or the isocurvature 
spectra. 

The spectra in these panels exhibit a large number of spike-like features 
that trace the intricate dynamics of the background trajectory. 
Sharp turns in field space enhance the derivative couplings appearing in 
\eqref{eq:m_orig}, which mix the different fluctuation mode components 
and transfer power between them. These transient mixing events propagate 
into the mode evolution and ultimately imprint themselves on the power 
spectra. These effects also propagate into the dynamics of the vielbein, 
governed by the transport equation \eqref{eq:transport_vielbein}, which 
maps the correlations from the parallel-transport gauge back to the 
original field-space basis. The solution of this additional system 
inherits the same rapid variations and therefore contributes nontrivially 
to the overall computational cost. Accurately resolving this dynamics 
requires retaining a sufficiently large number of modes and decreasing 
the evolution 
time-steps, the latter being partly driven by the need to capture the 
rapid variations of 
the background fields themselves. 

In contrast to the results shown in the upper panels, the nontrivial 
deformations $(\Delta V\neq 0)$ displayed in the lower panels of 
\Figref{fig:ps_k} require 
fewer modes (each corresponding to a different value of $k$) to be resolved. The associated power spectra exhibit markedly 
smoother features, which allows for larger evolution time-steps while 
maintaining numerical accuracy. This reduction in computational demand is 
not due to smaller potential deformations, but rather to the smoother variation of the 
background quantities and of the derivative couplings in 
\eqref{eq:m_orig}, which induce more gradual mixing among the different 
fluctuation mode components and therefore suppress the formation of sharp 
spectral structures. In addition to this, since the field-space metric 
remains flat in this case, no transport equations for the vielbeins need 
to be solved, further reducing the runtime. 

A further distinction concerns the background evolution itself: in contrast with 
the case of geometric deviations from flat field-space geometry, the total number 
of e-folds is sensitive to the magnitude of the potential deformations. Consequently, 
the end of inflation does not occur at a common location in field space across 
different deformation amplitudes. Evaluating the spectra at the end of inflation 
would obscure deformation effects, as differences accumulated during the evolution 
become entangled with variations in the termination point. To facilitate 
visualization, we instead place the collecting screen at a fixed number of e-folds, 
$N_{\rm e}$, enabling a direct comparison of spectral features. This choice is 
purely diagnostic and does not, in general, correspond to physically equivalent 
stages of the evolution.

\subsection{Sharp turns from potential deformations: models with two and more fields}
\label{subsec:res_potential}

Thus far, we have shown that the dynamical system proposed in 
\eqref{eq:EOM_AP_decomp_ReIm} efficiently captures the effects in the 
mode evolution arising from potential and geometric deformations sourced 
by arbitrarily sharp turns in the background field trajectory without 
introducing dynamical instabilities. In this subsection, we show that the 
same formalism characterizes how these sharp-turn effects are imprinted 
in the evolution of the covariance matrix, which parameterizes the Wigner 
quasiprobability distribution and, for purely Gaussian states, completely 
determines the corresponding density matrix operator. According to 
\eqref{eq:cov_r}, the covariance matrix is constructed solely from the 
amplitude variables and their derivatives; as described in 
\secref{sec:method}, therefore, its evolution inherits the symplectic 
structure that \eqref{eq:EOM_AP_decomp_ReIm} was designed to preserve. 
Together with the results of the previous subsection, this ensures that 
the framework can consistently solve the dynamics of the covariance 
matrix even in the presence of nonvanishing initial background field 
velocities. We further demonstrate that this amplitude-phase 
decomposition approach extends straightforwardly to inflationary models 
with more than two fields, retaining its structural and computational 
advantages in higher-dimensional field spaces.

\ellipses

To illustrate the impact of sharp turns on the evolution of the 
covariance matrix, we present in \Figref{fig:ellipses} the results 
obtained from the potential defined in \eqref{eq:potential_st}. It is 
important to recall that this parameterization was constructed to 
implement a continuous deformation of an equal-mass quadratic potential. 
In the configuration considered here, the deformation takes the form of 
helical modulations, which induce pronounced bends in the background 
field trajectory. The potential parameters used to generate these results 
are $m=1.4\times 10^{-6}M_{\rm Pl}$, $\lambda_{\rm V}=2.07\times 10^{-3}M_{\rm Pl}$, 
$f_1=12$, $f_2=0.5$ and $f_3=5.0M^{-1}_{\rm Pl}$. 

The left panel of \Figref{fig:ellipses} compares three 
representative cases with progressively increasing abruptness 
in the trajectory evolution. The solid green curve corresponds to the 
undeformed quadratic potential with vanishing initial field velocity, 
yielding a trajectory smoothly falling toward the center of the 
potential. The red curve retains zero initial velocity but includes 
nontrivial potential deformations, leading to noticeable turns. Finally, 
the yellow curve incorporates both nontrivial initial velocity and 
potential deformations, producing the sharpest and most abrupt trajectory 
features. In this case, the initial kinetic energy is sufficient for the 
fields to climb over the first potential barrier and subsequently 
encounter the next modulation, triggering the sharp turn visible in 
the figure. The dashed yellow line indicates that the evolution of the 
fluctuation mode begins half an e-fold after the black star, which denotes 
the starting point of the background trajectory.

The right panel of \Figref{fig:ellipses} presents a triangular plot 
representing a snapshot of the quantum state in terms of the canonical 
variables, combining phase-space projections and marginalized probability 
densities at a fixed instant of time. The off-diagonal panels display the 
Wigner ellipsoid projections in the corresponding two-dimensional phase-space 
planes at that instant, while the diagonal panels show the 
associated Gaussian probability density functions for each 
individual canonical variable. Each projected ellipse is constructed from the 
eigenvalues and eigenvectors of the corresponding two-dimensional covariance 
submatrix, with the eigenvectors determining its principal directions and the 
square roots of the eigenvalues determining its semiaxis lengths, with the contour 
normalized to unit Mahalanobis radius. We adopt the notation $p_X$ to denote the 
marginalized probability distribution of the canonical variable $X$. The three 
configurations correspond to those shown in the left panel and follow the same 
color coding. The plot captures the combined effects of the 
underlying inflationary expansion and the sharp turns, the latter entering 
through the derivative couplings in \eqref{eq:m_orig}. Despite this nontrivial 
dynamics, linearity and unitarity of the mode equations constrain the 
deformation of the Wigner ellipsoid to symplectic transformations. 
In particular, the phase-space volume enclosed by the ellipsoid is conserved, as 
explicitly demonstrated in \eqref{eq:cons_area} within the formulation of the 
amplitude-phase decomposition method developed in \subsecref{subsec:two_pt_pow_spec}. 
Consequently, the instantaneous geometry of the ellipsoid is entirely 
characterized by squeezing and phase-space rotations. Although the discussion 
presented here focuses on Gaussian states, the evolution of their phase-space 
geometry provides a useful diagnostic of the underlying dynamics. Strong 
squeezing and rapid rotations induced by structures in the potential (or in the 
field-space geometry) can 
signal regimes where nonlinear interactions become important, potentially 
leading to the generation of primordial non-Gaussianities. For 
completeness, the manuscript includes two ancillary animations.
The first presents the time-evolving counterpart of \Figref{fig:ellipses}, 
showing the continuous deformation of the Wigner ellipses and the associated 
marginalized distributions. The second illustrates the evolution of the 
ellipses in the presence of abrupt deformations in the field-space geometry 
parameterized according to \eqref{eq:fld_met_ex}\footnote{It is essential to 
clarify that the term geometric deformations refers exclusively to departures 
from a flat field-space metric. These should be distinguished from the 
continuous deformations of the potential discussed above, which can generate 
sharp turns in the background trajectory even when the field-space metric is 
flat.}.

\sixfieldsps

As a next step, we demonstrate that the amplitude-phase decomposition method proposed in this 
paper extends naturally to systems with an arbitrary number of fields. To this end, we consider 
the elliptic potential
\begin{equation}
V(\varphi^1,\varphi^2,\cdots,\varphi^6) = \frac{1}{2} \sum^{6}_{B=1} m^2_B(\varphi^B)^2 \, ,
\label{eq:elliptic}
\end{equation}
which we previously studied in the case $N_{\rm f}=2$ in \Figref{fig:cholesky_problem} to 
illustrate the shortcomings of the transport equation approach of \cite{GalvezGhersi:2016wbu} 
in the limit of large mass ratios. Effects in the mode solution are sourced by the 
inflationary trajectory, which begins from a corner of the potential configuration space, 
in a way that the high eccentricity of the potential induces potential walls, from which the 
trajectories bounce and subsequently oscillate around the attractor. Therefore, considering 
the background field dynamics and following the conventions for the power spectra described in 
\subsecref{subsec:two_pt_pow_spec}, \Figref{fig:6fields_ps} presents the evolution of the 
adiabatic and isocurvature modes, together with their cross-correlations, for the elliptic 
potential in the case of an inflationary model with six fields. We impose a hierarchy of masses 
defined by $m_1=5\times 10^{-8}M_{\rm Pl}$ and $m_B=5^{B-1}m_1$, such that the largest mass is 
three orders of magnitude greater than the smallest. The net effect of this hierarchy on the 
background evolution is reminiscent of a cascade in which the most massive field directions 
relax first, followed sequentially by progressively lighter directions, similar to what occurs 
in cascade models of inflation \cite{Ashoorioon:2006wc, Ashoorioon:2008qr, DAmico:2012khf}. As in all the 
cases illustrated in this section, the evolution does not show any interruptions or 
discontinuities that would indicate numerical or dynamical instabilities.

The upper panels of \Figref{fig:6fields_ps} show the time evolution of the spectra for a fixed 
comoving wavenumber. It is evident that sharp transitions arise from the hierarchy of masses. 
In contrast to cases with low mass ratios, the evolution of the cross-correlation power 
exhibits transient effects as the more massive field components relax first and the background 
trajectory successively reorients toward less massive field directions. The left panel 
illustrates the evolution of an adiabatic mode transitioning from a more to a less massive 
direction, giving rise to the characteristic ladder-like pattern. In the upper panels, the 
signed individual cross-correlation spectra $\mathcal{P}_{\mathcal{R}\mathcal{S}_{a}}$ 
associated with the more massive directions are shown in progressively lighter tones of red, 
while the corresponding isocurvature contributions 
$\mathcal{P}_{\mathcal{S}_{a}\mathcal{S}_{a}}$ are shown in progressively lighter tones of 
green. The total cross-correlation and isocurvature power spectra, $\mathcal{P}_{\mathcal{R}\mathcal{S}}$ and $\mathcal{P}_{\mathcal{S}\mathcal{S}}$, are shown 
as thick dark red and dark green dashed lines, respectively, while only these total spectra 
are displayed in the lower panels. The oscillatory features in the individual 
cross-correlation spectra, visible in the upper-middle panel, arise during the transient 
phases following successive encounters of the background trajectory with the steep 
potential walls, which induce oscillatory motion as the massive field components relax. 
In particular, the oscillatory behavior of the signed slow-roll parameter 
$\eta_{\parallel}$, whose absolute value is shown in the middle panel of 
\Figref{fig:slow_roll_params} for analogous two-field elliptic models, provides a qualitative 
two-field analogue of the sequence of oscillatory events observed here in the individual 
cross-correlation spectra. In the six-field model, each event is associated with the 
relaxation of a different massive background-field component and the resulting reorientation 
of the trajectory. As for the right panel, it shows that contributions from the 
more massive directions in the isocurvature sector are increasingly suppressed and therefore 
subdominant in the total power. It is worth noting that no substantial differences in 
execution time are observed when comparing the two- and six-field realizations of this model. 
Moreover, we find that the computational cost associated with resolving the mode evolution 
is comparable to that required for solving the background dynamics.

The lower panels of \Figref{fig:6fields_ps} illustrate how these dynamical features propagate
into the $k$-dependent spectra. Each localized bump can be traced to a step in the mode evolution, 
visible in the upper panels, arising as the more massive field components relax sequentially and 
the background trajectory is progressively redirected toward lighter field directions. In contrast 
to the upper panels, the corresponding contributions from the more massive directions to the final 
$k$-dependent spectra are strongly suppressed, to the point of being visually negligible. Notably, 
although the model considered here already demonstrates that the amplitude-phase decomposition 
method remains applicable in settings with more than two fields, we do not observe the emergence 
of distinctive features in the high-frequency regime of the spectra, suggesting that, within the 
class of potentials explored, the rapid-turn dynamics primarily imprint structure at low (or
intermediate) scales of $k$. While this behavior appears to be specific to the present setup, the
method is not restricted to such regimes and can be straightforwardly applied to scenarios in
which sharp turns and pronounced mass hierarchies do source localized features at smaller length
scales. This is particularly relevant in situations where enhanced curvature perturbations may
lead to the production of primordial black holes (see, \eg \cite{Wang:2024vfv}), as can occur
in inflationary models with nonminimal couplings to gravity
\cite{Garcia-Bellido:2011kqb, Pi:2017gih}, where these effects are known to induce amplification
at higher wavenumbers.

\subsection{Sharp turns driven by geometric deformations}
\label{subsec:res_geom}

The amplitude-phase decomposition method presented in this manuscript enables the
computation of all primordial power spectra and the full set of correlations 
between phase-space variables. The formulation introduced in 
\subsecref{subsec:eq_motion} naturally incorporates nontrivial 
field-space geometries through the inclusion of vielbein parallel transport 
equations in \eqref{eq:transport_vielbein}, providing a general framework to 
analyze geometry-induced effects in multifield dynamics. 

As an application, we consider the field-space geometry introduced in the 
geometrical destabilization scenario of \cite{Renaux-Petel:2015mga}, which 
induces nontrivial dynamical regimes in the background evolution. 
We combine this geometry with a potential that induces rapid turns in the 
background trajectory, providing a controlled environment in which geometry- and 
potential-induced effects contribute simultaneously to the dynamics, leading to 
distinctive features in the primordial spectra. These results are complementary 
to the geometric deformation considered in \eqref{eq:fld_met_ex}, shown in 
Figures \ref{fig:mode_evol} and \ref{fig:ps_k}. In practice, we restrict to 
values of $M$ for which the geometric deformation produces identifiable features 
in the spectrum, distinct from those induced by the potential, 
allowing the role of the field-space geometry to be unambiguously assessed.

As a first step, we recall the field-space geometry underlying the destabilization scenario in a
two-field model
\begin{align}
h_{AB}= \bigg[ 1 + \frac{2(\varphi^2)^2}{M^2} \bigg] \delta^1_A\delta^1_B+\delta^2_A\delta^2_B\,,
\label{eq:geom_dest_mt}
\end{align}
which, analogously to \eqref{eq:fld_met_ex}, can be expressed as a deformation of the flat 
field-space metric. For the potential, we consider the 
two-field version of the elliptic potential introduced at the end of the previous subsection, cf.~\eqref{eq:elliptic}, with a mass ratio of $m_2/m_1=15$. 

\renauxpetelmetric

With the field-space geometry and the potential defined, we show in 
\Figref{fig:renauxpetelmetric} the primordial curvature spectrum 
resulting from the combined impact of the elliptic potential in the large 
mass-ratio regime and the geometric deformation, with deviations from the 
flat-metric case controlled by the parameter $M$, as $M$ decreases. 
Darker (blue) curves correspond to configurations closer to the flat-metric 
regime, while lighter (yellow) curves indicate cases with stronger geometric 
deformations. As in \Figref{fig:ps_k}, we adopt identical initial conditions for 
all values of $M$ and evaluate the power spectrum after a fixed number of 
e-folds, set by the total expansion in the flat field-space case, $N_{\rm e}$, 
thereby providing a common reference time to directly compare the induced 
deformations. We emphasize that this choice does not correspond to a common 
physical end-of-inflation surface for all values of $M$; however, we have 
verified that the qualitative features described here persist when the spectra 
are evaluated at the end of inflation for each value of $M$, as illustrated in a
third ancillary animation included with this work. 

The large mass ratio in the 
elliptic potential induces a sharp transition in the spectrum, which evolves into
a pronounced peak as the geometric deformation increases. In this specific case, 
we find that the potential wall generated by 
the elliptic potential is sufficient to drive 
the slow-roll parameter, $\epsilon$, to values 
of $\mathcal{O}(1)$. Additionally, the effective squared mass for the isocurvature mode $\mathcal{M}^2_{\cal SS}\equiv\mathcal{M}^2_{AB}e^A_{\cal S}e^B_{\cal S}$ becomes negative in a narrow time interval for $M/M_{\rm Pl}\lesssim 4$. 
This behavior is consistent with 
the onset of geometrical 
destabilization and with the 
yellow and light green 
curvature peaks in 
the figure. 

\section{Discussion}
\label{sec:disc}

The purpose of this manuscript is to introduce a new framework for separating fast and slow 
dynamical scales in multifield inflation. While previous approaches, such as the Cholesky 
decomposition of correlators in \cite{GalvezGhersi:2016wbu}, perform this separation indirectly at 
the level of two-point functions, we instead reformulate the problem directly in terms of an 
amplitude-phase decomposition of the mode functions. This leads to a closed dynamical system 
that remains robust even in the presence of rapidly evolving background trajectories, such as 
those involving sharp turns. A key practical consequence of this formulation is that the evolution 
of the amplitude variables is governed by suppressed effective frequencies, allowing the mode 
dynamics to be accurately resolved using time steps comparable to those required for the 
background field evolution. Importantly, this construction is based on the preservation of the 
underlying symplectic structure of the system, and therefore maintains the equal-time canonical 
commutation relations throughout the evolution. This construction constitutes a natural 
generalization of the scale-separation approach developed for single-field models in 
\cite{GalvezGhersi:2018haa}, while incorporating the additional structure required for 
multifield dynamics. In particular, the use of parallel-transported vielbeins allows the method 
to consistently account for nontrivial field-space geometries, ensuring stable and accurate 
evolution across a broad class of multifield scenarios.

An additional advantage of the effective frequency suppression is that it allows modes to be 
consistently initialized deep inside the horizon, thereby opening the possibility of 
incorporating initial-state manipulation protocols and studying the dynamics of deformed Wigner 
ellipses in the presence of nontrivially evolving backgrounds. In \cite{Quispitupa:2025ayu}, we 
proposed a state-generation framework based on arbitrary sequences of decoherence and 
recoherence events 
\cite{Martin:2015qta,Martin:2021znx,Burgess:2022nwu}, which induce
controlled non-unitary deformations of the Wigner ellipsoid. After rescaling the conserved symplectic product, the output of such a deformation 
protocol can be directly used as an input state in the present framework, enabling the study of 
how these deformations evolve in backgrounds with sharp turns. In particular, the nontrivial 
dynamics associated with abrupt trajectory bending can propagate and amplify (or suppress) 
initial-state deformations through enhanced curvature-isocurvature mode mixing and 
cross-correlations. Importantly, the inclusion of such initial-state deformation protocols 
does not introduce a significant additional computational burden, as both the present approach 
and the state-deformation framework rely on similar strategies that suppress fast oscillatory 
scales and recast the dynamics in terms of slowly varying quantities.

A natural application of the present framework is the computation of primordial spectra in 
multifield models with random potentials 
\cite{Vafa:2005ui,Frazer:2011tg,Higaki:2014mwa,Linde:2016uec,Furuta:2025tjt}, as motivated by 
landscape (or swampland) considerations. In these scenarios, the underlying potentials are 
typically steep and highly structured, leading to background trajectories that exhibit frequent 
and rapid turns in field space, making them a well-suited arena for the methods developed here. 
At the same time, it is well known that achieving a sufficiently long period of inflation in 
genuinely random constructions is statistically rare, which motivates the construction of 
constrained random potentials that retain their stochastic features while being engineered to 
sustain prolonged inflation. In exploratory attempts to realize such scenarios, we find that 
unconstrained random potentials generically either fail to produce a sufficient number of 
e-folds or instead lead to regimes of eternal inflation. Regardless of the outcome, we find 
that the present approach provides a robust tool to analyze the resulting dynamics and to 
reliably compute the associated primordial 
spectra in these nontrivial settings. If 
realizations exhibiting sufficient inflation can be identified with reasonable frequency, 
the amplitude-phase decomposition framework enables efficient Monte-Carlo explorations, 
opening the possibility of placing observational bounds on the parameters defining each 
realization. A systematic construction of such constrained potentials from Gaussian random
fields will be developed in forthcoming work.


\acknowledgments

The authors would like to thank Guillem Domenech, Andrei Frolov, Stefano Gonzales, 
Lorena Luján, Jorge Medina, Diego Suárez and Sashwat Tanay for many fruitful discussions, 
as well as their technical support and feedback in the first versions of this paper. 
Part of the computations were performed on the cloud computational resources provided 
by Oracle. The work of GQ was 
funded in part by NSERC Discovery Grant ``Testing fundamental physics with B-modes on 
Cosmic Microwave Background anisotropy''. The work of JP and JG was partially funded by Fondo Semilla 2023, granted by Universidad de Ingeniería y Tecnología.
This project was funded by CONCYTEC through the PROCIENCIA program in the context of 
the research opportunity ``E041-2024-03. Proyectos de Investigación Básica'', 
contract number PE501087705-2024-PROCIENCIA.

\bibliography{Bibnotes.bib}

\end{document}